\documentclass[12pt]{article}
\usepackage{epsfig,amsfonts,amssymb}
\usepackage{amsmath}
\usepackage[margin=2.5cm]{geometry}
\usepackage[colorlinks=true,urlcolor=blue,linkcolor=blue]{hyperref}
\usepackage{cite}
\input epsf.sty
\usepackage{cite}

\def\ZZZ{{\hbox{ Z\kern-1.6mm Z}}}
\def\RRR{{\hbox{ R\kern-2.4mm R}}}
\def\CCC{{\hbox{ C\kern-2.0mm C}}}
\def\zzz{{\hbox{z\kern-1mm z}}}

\newcommand{\vt}{\vartheta}

\newcommand{\qeq}{{\hbox{=\kern-2.3mm ? \kern.5mm }}}
\renewcommand{\qeq}{=}

\newcommand{\eps}{\epsilon}

\newcommand{\vp}{\varphi}
\newcommand{\ve}{\varepsilon}

\newcommand{\MM}{{\cal M}}

\newcommand{\WW}{{\cal W}}

\newcommand{\wh}{\widehat}

\newcommand{\NN}{{\cal N}}

\newcommand{\be}{\begin{equation}}
\newcommand{\ee}{\end{equation}}
\newcommand{\ben}{\begin{eqnarray}\displaystyle}
\newcommand{\een}{\end{eqnarray}}

\newcommand{\refb}[1]{(\ref{#1})}
\newcommand{\p}{\partial}
\newcommand{\sectiono}[1]{\section{#1}\setcounter{equation}{0}}

\def\one{{\hbox{ 1\kern-.8mm l}}}
\def\zero{{\hbox{ 0\kern-1.5mm 0}}}

\newcommand{\Tr}{{\rm Tr}}

\def\nnn{\nonumber \\ && }

\begin{document}

\baselineskip 24pt

\begin{center}
{\Large \bf  Do All BPS Black Hole Microstates 
Carry Zero Angular Momentum?}

\end{center}

\vskip .6cm
\medskip

\vspace*{4.0ex}

\baselineskip=18pt

\centerline{\large \rm Abhishek Chowdhury$^a$, Richard S.~Garavuso$^b$, 
Swapnamay Mondal$^a$, Ashoke Sen$^a$}

\vspace*{4.0ex}

\centerline{\large \it ~$^a$Harish-Chandra Research Institute}
\centerline{\large \it  Chhatnag Road, Jhusi,
Allahabad 211019, India}

\centerline{\large \it ~$^b$Physical Sciences, 
Kingsborough Community College}
\centerline{\large \it The City University of New York}
\centerline{\large \it 
2001 Oriental Boulevard,
Brooklyn, NY 11235-2398, USA}

\vspace*{1.0ex}
\centerline{\small E-mail:  
abhishek,swapno,sen@hri.res.in,  richard.garavuso@kbcc.cuny.edu}

\vspace*{5.0ex}

\centerline{\bf Abstract} \bigskip

From the analysis of the near horizon geometry and supersymmetry
algebra it has been argued that all the microstates of single centered
BPS black holes with four unbroken supersymmetries
carry zero angular momentum in the region of the moduli space where
the black hole description is valid. A stronger form of the conjecture would
be that the result holds for any sufficiently generic point in the moduli
space. In this paper we set out to test this conjecture for a class of black
hole microstates in type II string theory on $T^6$, represented by four stacks
of D-branes wrapped on various cycles of
$T^6$. For this system the above conjecture translates to the statement
that the moduli space of classical vacua must be a collection of points.
Explicit analysis of systems carrying a low
number of D-branes supports this conjecture.

\vfill \eject

\baselineskip=18pt

\tableofcontents

\sectiono{Introduction}  \label{s1}

Based on the analysis of the near horizon geometry of a BPS black hole it has
been argued that as long as the black hole carries four unbroken supersymmetries,
the microstates of single centered black holes will all carry zero angular
momentum\cite{0903.1477,1009.3226}.\footnote{This statement 
requires some qualification since
all BPS black holes form a representation of the supersymmetry algebra carrying
many different spins. This arises from the quantization
of the fermion zero modes associated with broken supersymmetry.
In the black hole description these modes live {\it outside} the horizon. On the
other hand, the
property of zero angular momentum refers to the microstates associated with
the black hole horizon. Therefore the correct statement is that the microstates of
single centered BPS black holes are given by the tensor product of the elementary
BPS supermultiplet, obtained by quantization of the goldstino
fermion zero modes, and singlet representation of the SU(2) rotation group.
For black holes in five dimensions a further qualification is needed -- the black
hole microstates are singlets under one of the two $SU(2)$'s-- the one that forms
a subgroup of the $SU(1,1|2)$ supergroup describing symmetries of the near
horizon geometry. This complication will not affect our analysis since we
shall focus on black holes in 3+1 dimensions.} 
 Since the counting of black hole microstates is normally done in a regime that
 is different from the regime in which the black hole description is valid, 
 it is difficult to test this conjecture directly. Indirect tests of this have been
performed by examining the BPS index -- given by number of states weighted
by $(-1)^{2J_3}$ after factoring out the goldstino fermion zero mode 
contributions -- which is expected to be independent of
the moduli. On the black hole side if all the microstates carry zero angular momentum,
then the BPS index will be positive. This implies that on
the microscopic side also we must always have positive index. This has been
tested in many examples\cite{1008.4209,1208.3476}, and so 
far no counterexample has been found, even
though the microscopic counting typically gives states carrying different
angular momenta (even 
after factoring out the the goldstino zero mode contribution), some of which
give positive contribution to the index and others give negative contribution.

If this conjecture is correct, then one might wonder at what point in the moduli
space does the spectrum of BPS states change from having various different
angular momenta to only zero angular momentum?  A natural conjecture 
-- which we shall call the strong form of the zero angular momentum conjecture --
will
be that this happens at any sufficiently generic point in the moduli space, even
in regions where the system is weakly coupled and the microscopic description
is valid. This is not in direct conflict with the results for the microscopic
spectrum described at the end of the previous paragraph, 
since typically the microscopic spectrum is computed not only
by setting the string coupling to be weak, but also by setting several other
moduli to special values where the quantum mechanics describing the microstates
simplifies. 

A possible complication in testing this conjecture is that it refers to the
microstates of single centered black holes. Even in regions where the black
hole description is valid, typically there are multi-centered BPS black hole
configurations contributing to this index, and these typically carry non-zero
angular momentum. In $\NN=2$ supersymmetric theories in 3+1 dimensions
the contribution from the multi-centered black holes is very 
complicated\cite{0702146}
and it is not easy to disentangle it from the contribution from single centered
black holes, although some progress has been made on this 
front\cite{1103.1887}.  In this context the results 
of \cite{1205.5023,1205.6511,1207.0821,1207.2230,1302.5498}
can be taken as providing partial support to this conjecture.
The situation is somewhat better in $\NN=4$ supersymmetric theories
where the contribution from multi-centered black holes are better 
understood\cite{0702141,0702150,0705.3874,0706.2363}.
But the situation is best for $\NN=8$ supersymmetric theories, where for the
charge configuration of the kind for which we have black holes, there is
no contribution to the index from multi-centered black 
holes\cite{0803.1014} and in fact all 
multi-centered BPS states are expected to disappear at a generic point in the
moduli space. Therefore in this case the strong form of the conjecture will imply
that at a generic point in the moduli space, all the microstates must carry
zero angular momentum.

In a previous paper\cite{1405.0412} 
we began to test this conjecture by considering a system
of four stacks of D-branes in type II string theory of 
$T^6$. If we denote by 
$x^4,\ldots, x^9$ the coordinates along $T^6$, then in an appropriate
duality frame the general system considered in \cite{1405.0412} 
can be represented as
$N_1$ D2-branes wrapped along the 4-5 directions, $N_2$ D2-branes wrapped
along the 6-7 directions, $N_3$ D2-branes wrapped along the 8-9 directions and
$N_4$ D6-branes wrapped along the 4-5-6-7-8-9 directions. 
We keep the moduli sufficiently generic by switching on generic values
of the constant metric and NS-NS 2-form background along $T^6$.
The low energy dynamics of
the system was shown to describe a specific quantum mechanical system, obtained
from the dimensional reduction of an $\NN=1$ supersymmetric field theory in
3+1 dimensions to 0+1 dimensions. The moduli space describing the classical
supersymmetric solutions\footnote{This 
should not be confused with the moduli space
describing background closed string fields which are kept at generic values.}
corresponds to gauge inequivalent solutions to the F- and D-term constraints of the theory, and the quantum BPS states are 
given by harmonic forms on this moduli space\cite{Witten:1982df}. 
By considering the simplest 
system in which we have one D-brane in each of the four stacks, it was found
that except for trivial flat directions associated with translation along the
non-compact directions as well as along $T^6$ and
dual $T^6$, the moduli space is a collection of 12 points. The quantization of the
translational zero modes produces states carrying different momenta along the
non-compact directions and along $T^6$ and also different fundamental string
winding charges along $T^6$. Once these quantum numbers are fixed we
get a unique state from quantization of the bosonic zero modes. 
The fermionic partners of these bosonic zero modes
are the 28 goldstino fermion zero modes, whose quantization produces the
basic BPS supermultiplet. The fact that  after factoring out
the zero mode contributions the moduli space becomes a collection of
isolated points shows that
the BPS ground state associated with each solution is unique, and hence
must be singlets of the $SU(2)$ rotation group.\footnote{A similar result for the
D0-D4 system was found in \cite{0001189}.}
The 12 isolated configurations
then lead to 12 BPS states, each carrying zero angular momentum.
This is consistent with the
strong conjecture stated above. This also shows that the BPS index of this
system is 12. This is in perfect agreement with the result of the computation
of the index for the same system in a dual 
description\cite{9903163,0506151}.

In this paper we extend the analysis of \cite{1405.0412} to the 
cases where the number of 
D6-branes in the stack is 2 or 3. Again we find that for generic constant
background values
of metric and NS-NS 2-form fields, the moduli space of classical supersymmetric
configurations consists of a discrete set of points after factoring out
the zero mode contribution. This in turn shows that after 
factoring out the basic BPS supermultiplet we are left with a set of singlet
representations of the $SU(2)$ rotation group: 56 for the case of 2 D6-branes
and 208 for the case of 3 D6-branes. This is again consistent with the strong
form of the conjecture stated above. Furthermore since all states are
SU(2) singlets, the BPS indices associated with these states are also
given by 56 and 208, respectively. This agrees precisely with the index
computed in a dual description\cite{0506151}. However in \cite{0506151}
the counting was carried out at a non-generic point in the moduli space, and
the spectrum contained BPS states carrying different angular momenta. 
The
index was the final result after imperfect cancellation between contributions
from bosonic and fermionic states. In contrast here all states carry zero
angular momentum and give positive contribution to the index.

The rest of the paper is organized as follows. In \S\ref{ssystem} we describe
the system under study, and review the results of \cite{1405.0412} 
for the supersymmetric
quantum mechanics describing the low energy dynamics of this system.
In \S\ref{s2.5} we describe in detail the steps we follow for counting the number
of BPS states consisting of a system of 
one D2-brane along 4-5 directions, one D2-brane along 6-7 directions, one
D2-brane along 8-9 directions and one
D6-brane along 4-5-6-7-8-9 directions.
In \S\ref{s3} we describe the analysis for 
a system consisting of one D2-brane along 4-5 directions, one D2-brane along 
6-7 directions, one
D2-brane along 8-9 directions and two
D6-branes along 4-5-6-7-8-9 directions. In \S\ref{s4} we briefly describe the
analysis for three D6-branes along 4-5-6-7-8-9 directions, leaving the number of
D2-branes unchanged. We conclude in \S\ref{sconc} with a discussion
on the possible implications of our results in general, and in particular for the
fuzzball program.
In appendix \ref{sC} we carefully fix the signs and
normalizations of various terms in the superpotential -- the analysis of 
\cite{1405.0412} was insensitive to these choices but they seem to be relevant
for analyzing the general case with multiple D-branes in one or more
stacks. In this we fix the normalization by explicit string theory computation and
the signs by various symmetry requirements. We are however left with a 
twofold ambiguity in the choice of sign, only one of which yields the correct
answer in different cases we have studied. While  a careful string theory
computation should be able to resolve this ambiguity, we have not done this.
In appendix \ref{ea1} we give explicit results for the 56 solutions in the special case
when one of the stacks contains two D-branes with the other stacks containing one
D-brane each.

\sectiono{The system} \label{ssystem}

We consider type IIA string theory on $T^6$ labelled by the coordinates
$x^4,\ldots, x^9$ and in this theory we take a system
containing one D2-brane wrapped along
4-5 directions, one D2-brane wrapped along 6-7 directions, one 
D2-brane wrapped
along 8-9 directions and $N$
D6-branes wrapped along 4-5-6-7-8-9 directions, where $N$ takes
values 1, 2 and 3.\footnote{Alternatively, 
by making T-duality transformations along each of the six circles
we can regard this as a system of one D4-brane along each of the three
4-cycles  4567, 6789 and 4589, and $N$ D0-branes. Under this transformation
the quantum theory will remain the same, but the physical
interpretation of the different variables given below will differ, with positions
and Wilson lines getting exchanged. 
 \label{f1}}
As in \cite{1405.0412} we shall
work at a generic point in the moduli space at which constant metric and
2-form backgrounds are turned on along $T^6$.
The supersymmetric quantum mechanics describing this system was 
constructed in \cite{1405.0412}. Since this system has four unbroken supercharges
-- the same as for $\NN=1$ supersymmetric theories in four dimensions -- we shall
use the language of four dimensional theories to organize the fields and terms
in the Lagrangian. However since we are really considering a quantum mechanical
system, we ignore all spatial derivative terms in the action. 

We begin by listing the field content of this 0+1 dimensional field
theory. We shall focus only on the bosons since
the fermionic field content is determined by supersymmetry.  The four dimensional
theory can be regarded as an $\NN=1$ supersymmetric theory with gauge group
$U(1)\times U(1)\times U(1)\times U(N)$ with additional chiral multiplets.
We denote by $X_i^{(k)}$ for $1\le k\le 4$, $1\le i\le 3$ the $i$-th spatial
component of these four sets of gauge fields. $k=1,2,3$ will stand for the U(1)
factors and $k=4$ will stand for the $U(N)$ factor.\footnote{There are also non-dynamical 
fields $X^{(k)}_0$ which implement the gauge invariance constraints. Their effect will be
discussed later.}
Therefore $X^{(4)}_i$ denotes an
$N\times N$ Hermitian matrix for each $i$ whereas 
$X^{(k)}_i$ for $1\le k\le 3$, $1\le i\le 3$
are real numbers. Physically the $X^{(k)}_i$ for $1\le k\le 3$ describe the  position
of the $k$-th D2-brane along the three non-compact directions and the diagonal elements
of $X^{(4)}_i$ label the
positions of the $N$ D6-branes along the non-compact directions. 
The off-diagonal components
of $X^{(4)}_i$ arise from open strings stretched between the D6-branes.
The rest of the bosonic fields can be organized into chiral multiplet
scalars. First of all, for each of the four stacks
of D-branes
we have three chiral multiplets in the adjoint representation. 
We shall denote them by $\Phi^{(k)}_i$ for $1\le k\le 4$,
$1\le i\le 3$. $\Phi^{(k)}_i$ for $1\le k\le 3$ describe $1\times 1$ complex matrices whereas
$\Phi^{(4)}_i$ will describe an $N\times N$ complex matrix, transforming in the adjoint representation 
of the respective gauge groups. 
Physically the $\Phi^{(k)}_i$'s for $1\le k\le 3$  label the  position of the $k$-th
D2-brane along the directions of $T^6$ transverse to the D2-brane, and also the Wilson lines 
on the $k$-th D2-brane along directions tangential to the D2-brane. On the other hand the diagonal
elements of $\Phi^{(4)}_i$
label the Wilson lines along the D6-branes. 
The off-diagonal components
of $\Phi^{(4)}_i$ arise from open strings stretched between the D6-branes.
Besides these fields, for every pair of D-brane stacks labelled by
$(k,\ell)$ we have two 
chiral superfields $Z^{(k\ell)}$ and
$Z^{(\ell k)}$ arising from open strings stretched between the two D-brane stacks, transforming respectively in the
bi-fundamental and anti-bi-fundamental representation of the corresponding gauge groups. 
Therefore $Z^{(k\ell)}$ for $1\le k,\ell\le 3$ are $1\times 1$ matrices, $Z^{(k4)}$ for $1\le k\le 3$ are
$1\times N$ matrices and $Z^{(4k)}$ for $1\le k\le 3$ are $N\times 1$ matrices.  Under the rotation
group $SU(2)$ acting on the non-compact directions, $X^{(k)}_i$'s transform as vectors whereas
$\Phi^{(k)}_i$'s and $Z^{(k\ell)}$'s transform as scalars.

The potential involving these fields receives contributions from several sources. First of all
we have several terms in the superpotential.\footnote{The superpotential may contain
other terms, {e.g.} terms quartic in $Z^{(k\ell)}$'s. However if we consider the
limit when all the $c^{(k\ell)}$'s and $c^{(k)}$'s 
are small, say of order $\lambda$ for some small
parameter $\lambda$, then the solutions to the F- and D-term 
equations occur at
$Z^{(k\ell)}, \Phi^{(k)}_i\sim \sqrt{\lambda}$. In this limit the effect of the other terms
in the superpotential can be ignored. This limit also allows us to ignore the fact that the
$\Phi^{(k)}_i$ fields have periodic identification.
\label{f77}}
They are given by\cite{1405.0412}
\ben \label{ew1gen}
\WW_1 &=& \sqrt 2\left[\sum_{k,\ell,m=1}^3 \ve^{k\ell m} \Tr \, \Big(\Phi^{(k)}_m
\, Z^{(k\ell)} Z^{(\ell k)} \Big) + \sum_{k=1}^3 \Tr \, \Big(  Z^{(4k)} \Phi^{(k)}_k Z^{(k 4)} \Big)\right.
\nnn\left.
\qquad - \sum_{k=1}^3 \Tr \, \Big(   \Phi^{(4)}_k Z^{(4k)} Z^{(k 4)} \Big)\right] \nonumber \\
&=& \sqrt 2 \bigg[ (\Phi^{(1)}_3 - \Phi^{(2)}_3) Z^{(21)} Z^{(12)} + 
(\Phi^{(2)}_1 - \Phi^{(3)}_1) Z^{(32)} Z^{(23)} + (\Phi^{(3)}_2 - \Phi^{(1)}_2) Z^{(31)} Z^{(13)} \nonumber \\ &&
+ {\rm Tr} \Big((\Phi^{(1)}_1 I_N - \Phi^{(4)}_1) Z^{(41)} Z^{(14)} \Big)
+ {\rm Tr} \Big((\Phi^{(2)}_2 I_N - \Phi^{(4)}_2) Z^{(42)} Z^{(24)} \Big) \nonumber \\ &&
+ {\rm Tr} \Big((\Phi^{(3)}_3 I_N - \Phi^{(4)}_3) Z^{(43)} Z^{(34)} \Big)
\bigg]
\, ,
\een
\ben \label{ew2gen}
\WW_2 &=& \sqrt 2\,
\left[\sum_{k,\ell, m=1\atop k<\ell,m; \, \ell\ne m}^4 (-1)^{\delta_{k1} \delta_{\ell 3} \delta_{m4}} 
\Tr \Big(Z^{(k\ell)} Z^{(\ell m)} Z^{(m k)}
\Big)\right]  \nonumber \\ &=&  \sqrt 2\,  \bigg[
Z^{(31)} Z^{(12)}Z^{(23)} + Z^{(13)} Z^{(32)}Z^{(21)}  + {\rm Tr} \Big( Z^{(12)} Z^{(24)}Z^{(41)}\Big)
+ {\rm Tr} \Big( Z^{(42)} Z^{(21)}Z^{(14)} \Big) \nonumber \\ &&
- {\rm Tr} \Big( Z^{(13)} Z^{(34)}Z^{(41)}\Big)
+ {\rm Tr} \Big( Z^{(31)} Z^{(14)}  Z^{(43)}\Big)+ {\rm Tr} \Big( Z^{(34)} Z^{(42)}Z^{(23)} \Big)
\nonumber \\ && + {\rm Tr} \Big( Z^{(43)} Z^{(32)}Z^{(24)} \Big)
\bigg]
\, ,
\een
\ben \label{ew3gen}
\WW_3 &=& \sqrt 2\left[\sum_{k,\ell,m=1}^3 c^{(k\ell)} \, \ve^{k\ell m}  \, 
\Tr \, \Big(\Phi^{(k)}_m \Big) 
+ \sum_{k=1}^3 c^{(k4)} \, \Big[ N_4\, 
\Tr \, \Big(\Phi^{(k)}_k\Big)  - \Tr\Big( \Phi^{(4)}_k\Big) \Big]\right]\nonumber \\ 
&=& \sqrt 2\, 
\bigg[c^{(12)} (\Phi^{(1)}_3 - \Phi^{(2)}_3) + c^{(23)} (\Phi^{(2)}_1 - \Phi^{(3)}_1) 
+ c^{(13)} (\Phi^{(3)}_2 - \Phi^{(1)}_2) + c^{(14)} \Tr \Big(\Phi^{(1)}_1 \, I_N - \Phi^{(4)}_1\Big) 
\nonumber \\ &&
+ c^{(24)} \Tr \Big(\Phi^{(2)}_2 \, I_N - \Phi^{(4)}_2\Big) + c^{(34)} \Tr \Big(\Phi^{(3)}_3 \, I_N 
- \Phi^{(4)}_3\Big) 
\bigg]
\, ,
\een
and
\be \label{ew4gen}
\WW_4 = -\sqrt 2 \, \Tr \Big( \Phi^{(4)}_1 \left[\Phi^{(4)}_2, \Phi^{(4)}_3\right] \Big)
= - \sqrt 2 \, \Tr \Big( \Phi^{(4)}_1\Phi^{(4)}_2\Phi^{(4)}_3 - \Phi^{(4)}_1\Phi^{(4)}_3\Phi^{(4)}_2\Big) \, ,
\ee
where $c^{(k\ell)}=c^{(\ell k)}$ for $1\le k<\ell\le 4$ are parameters whose values
are determined in terms of the background metric and 
2-form fields\cite{1405.0412}, and
$I_{p}$ denotes the
$p\times p$ identity matrix.
There are some differences from the form of the superpotential given in 
\cite{1405.0412}. First of all in \cite{1405.0412} the overall coefficient of 
$\WW_2$ was undetermined. For $N=1$ 
this constant can be scaled away by appropriate scaling of the fields $Z^{(k\ell)}$,
$\Phi^{(k)}_i$ and the parameters $c^{(k\ell)}$, and so 
its value
was not needed. However for $N\ge 2$ this can no longer be done without
changing the coefficient of the cubic $\Phi$-$\Phi$-$\Phi$ coupling.
Since the result depends on the choice of
this coefficient, we have fixed it from explicit string theory
computation described in appendix \ref{sC}. 
Second, we have a 
strange sign $(-1)^{\delta_{k1} \delta_{\ell 3} \delta_{m4}}$ in the expression for
$\WW_2$ which is responsible for the minus sign in front of 
the ${\rm Tr} \Big( Z^{(13)} Z^{(34)}Z^{(41)}\Big)$ term. 
This sign was missed in \cite{1405.0412} but must be there for symmetry 
reasons. This has also been explained in appendix \ref{sC}.\footnote{As discussed in \S\ref{s2.5},
the result for the number of states for $N=1$, 
obtained in \cite{1405.0412}, is not affected by
this change of sign.}
In
appendix \ref{sC} we also discuss generalization of \refb{ew1gen}-\refb{ew4gen}
to the case where we have an arbitrary number of D-branes in each of the
four stacks.

The F-term potential is given by
\be \label{efterm}
V_F=\sum_\alpha \left|{\p \WW\over \p\vp_\alpha}\right|^2\, ,
\ee
where 
\be
\WW=\WW_1+\WW_2+\WW_3+\WW_4\, ,
\ee
and $\{\vp_\alpha\}$ denotes the set of all the chiral
superfields.

Besides the $F$-term potential,
there is a D-term potential given by
\be \label{evdgen}
V_D = {1\over 2} \, 
\sum_{k=1}^4   \Tr \bigg[\Big( \sum_{\ell=1\atop \ell \ne k}^4 Z^{(k\ell)} Z^{(k\ell)\dagger} 
- \sum_{\ell=1\atop \ell \ne k}^4 Z^{(\ell k)\dagger}  Z^{(\ell k)}  + 
\sum_{i=1}^3  [\Phi^{(k)}_i, \Phi^{(k)\dagger}_i] -  c^{(k)} I_{N_k} \Big)^2 \bigg]\, ,
\ee
where $N_1=N_2=N_3=1$ and $N_4=N$. 
The FI parameters
$c^{(k)}$  are also determined from the background values of the 2-form field and
satisfy
\be \label{ecknonabelian}
\sum_{k=1}^4 c^{(k)} N_k = 0\, .
\ee

Finally the dimensional reduction of the coupling of the gauge fields 
(whose spatial components are denoted by $X^{(k)}_i$) to chiral multiplets
leads to the potential
\ben \label{egaugegen}
V_{gauge} &=&\sum_{k=1}^4 \sum_{\ell =1\atop \ell \ne k}^4  \sum_{i=1}^3   \,
\Tr \Big[\Big( X^{(k)}_i  Z^{(k\ell)} - Z^{( k\ell)} 
X^{(\ell)}_i \Big)^\dagger  \Big( X^{(k)}_i  Z^{(k\ell)} - Z^{( k\ell)} 
X^{(\ell)}_i \Big)\Big] \nonumber \\
&& + \sum_{k=1}^4 \sum_{i,j=1}^3 \Tr \Big(\big[X^{(k)}_i, \Phi^{(k)}_j\big]^\dagger
\big[X^{(k)}_i, \Phi^{(k)}_j\big]\Big) + {1\over 4} \sum_{k=1}^4 \sum_{i,j=1}^3 
\Tr \Big( [X^{(k)}_i, X^{(k)}_j]^\dagger [X^{(k)}_i, X^{(k)}_j]\Big)\, . \nonumber \\
\een
Therefore the total potential is given by
\be \label{evtotal}
V = V_F+V_D+V_{gauge}\, .
\ee

The potential given above has a shift symmetry
\ben \label{eflatgen}
&& \Phi^{(k)}_m \to \Phi^{(k)}_m+\xi_m,  \quad \hbox{for} 
\quad 1\le k \le 3, \quad k \ne m, \quad  1\le m\le 3, 
\nonumber \\
&& \Phi^{(k)}_k \to \Phi^{(k)}_k + \zeta_k, \quad \Phi^{(4)}_k \to
\Phi^{(4)}_k+\zeta_k I_{N}, \quad \hbox{for} \quad 1\le k\le 3\, , \nonumber \\
&& X^{(k)}_i \to X^{(k)}_i + a_i, \quad X^{(4)}_i \to X^{(4)}_i + a_i\,  I_N, \quad \hbox{for} \quad 1\le i\le 3,
\quad 1\le k\le 3\, ,
\een
where $\{\xi_m\}$ and $\{\zeta_k\}$ are arbitrary complex parameters
and $\{a_i\}$ are arbitrary real parameters.
These shift symmetries
generate six complex translations along the compact directions 
and their duals and three real translations
along the non-compact directions. 

Our strategy for determining the spectrum of BPS states will be to regard the low
energy dynamics of the system as the motion of a superparticle moving on the
classical vacuum manifold defined by the space of $V=0$ configurations, and
then quantize the system and find its supersymmetric ground states. Now the shift
symmetries \refb{eflatgen} generate flat directions of the potential $V$.
Quantization of the bosonic zero modes 
associated with these flat directions leads to 
a unique ground
state if we restrict to the sector carrying zero momentum and winding 
along the internal directions and zero momentum along the non-compact
directions. There are also associated
fermionic zero modes describing the goldstino modes corresponding to $32-4=28$ 
broken supersymmetries. Quantization of these fermion zero modes produces the supermultiplet
describing 1/8 BPS states of $\NN=8$ supersymmetric string theory but has no other effect
on the rest of the system. 
The BPS spectrum is given by a tensor product of this basic supermultiplet
with some (possibly reducible) representation of the rotation group $SU(2)$.
Our goal will be to determine which representation of $SU(2)$ is tensored with the
basic supermultiplet.

The information
about the
SU(2) representation with which the supermultiplet is tensored
is contained in the character $ P (y) \equiv Tr(y^{2 J_3})$ of the representation.
$ P (y)$ is computed as follows. 
We shall show at the end of appendix \ref{sC} 
that the vanishing of $V_{gauge}$ given in \refb{egaugegen}
requires all the $X^{(k)}_i$'s to vanish (up to the shift symmetry described in the
last line of \refb{eflatgen}). 
If the gauge inequivalent solutions to the $V_F=V_D=0$ condition generate a 
manifold $\MM$ of complex dimension $d$ after factoring out the flat directions
of the potential
associated with the symmetries given in \refb{eflatgen}, and setting the $X^{(k)}_i$'s
to zero,
then the BPS states of the system are in one to one correspondence with
the harmonic forms on $\MM$, and the rotational $SU(2)$ is identified with the
Lefschetz SU(2) acting on these 
forms (see {\it e.g. \cite{0206072,1205.5023}}).
Therefore if $b_p$ denotes the $p$-th betti
number of $\MM$ and $d$ denotes the complex dimension of $\MM$, 
then we have\footnote{Intuitively this 
identification can be
understood as follows. 
Since the vacuum manifold has $X^{(k)}_i=0$
for all $k,i$, the moduli space is spanned by the scalars. The fermionic
partners of the scalars take values in the tangent space of $\MM$ 
-- in fact for each tangent
vector there are two massless fermions 
which we can denote by $\psi^a$ and 
$\psi^{a\dagger}$ where $a$ labels independent tangent vectors. We can choose
$\psi^a$ and 
$\psi^{a\dagger}$ such that $\psi^{a\dagger}$ has $J_3=1/2$ and 
$\psi^a$ has $J_3=-1/2$. Now we can begin with the states
annihilated by all the $\psi^a$'s, identify them as the zero forms on $\MM$, and
build the total space of states by applying  $\psi^{a\dagger}$'s on this state.
This space is isomorphic to the space of forms on $\MM$, and
the BPS condition translates to these forms being harmonic. In this notation
we see that the $p$-forms carry $J_3$ eigenvalue $(p-d)/2$, where the shift 
$-d/2$ is the $J_3$ eigenvalue of the zero forms, and is necessary to ensure
that the states form a representation of $SU(2)$. This leads to  \refb{espectrum}.
}
\be \label{espectrum}
 P (y) = \sum_p b_p y^{p-d}\, .
\ee
If $\MM$ contains several components then we have to add up the contribution from
various components to get the total $ P (y)$. 
The BPS index (which is the 14-th helicity trace $-B_{14}$ 
\cite{9611205,9708062} from the space-time
viewpoint) is given by $Tr(-1)^{2J_3}$, with the trace
running over the states with which the basic BPS supermultiplet is tensored to get
the full spectrum of BPS states. Therefore it is given by $ P (-1)$, which, according
to \refb{espectrum}, is $(-1)^d$ times  the Euler character of $\MM$.

The conjecture that all BPS states carry zero angular momentum now translates
to the requirement that $ P (y)$ is $y$-independent, i.e. the subspace $\MM$
consists of isolated points.
In this case the BPS index $ P (-1)$
is equal to the degeneracy $ P (1)$ and just counts the number of gauge 
inequivalent solutions to the $V=0$ condition.

We conclude this section by 
giving the expected result for the index from analysis
of the spectrum in a dual description containing a D1-D5-KK monopole 
state carrying
momentum along the common circle shared by the world-volume of the D1-brane,
D5-brane and the KK monopole. Let us 
denote by $\hat c(u)$ the numbers appearing in the
expansion
\be \label{ek6.5}
-\vt_1(z|\tau)^2 \, \eta(\tau)^{-6} \equiv \sum_{k,l} \hat c(4k-l^2)\, 
e^{2\pi i (k\tau+l z)}\, ,
\ee
where $\vt_1(z|\tau)$ and $\eta(\tau)$ are respectively the odd Jacobi
theta function and the Dedekind eta function. Then the expected result for
the index $ P (-1)$ is\cite{0506151}\footnote{Various 
macroscopic tests of this formula beyond the one provided by the
Bekenstein-Hawking formula
have been carried out in \cite{1005.3044,1106.0080,
1012.0265,1111.1161,1404.0033}. \label{f6}}
\be \label{epr1}
- \hat c(4N)\, .
\ee
Explicit computation gives
\be  \label{epr2}
-\hat c(4)=12, \quad -\hat c(8) =56, \quad -\hat c(12) = 208, \quad -\hat c(16)=684,
\quad \cdots
\ee
In the previous paper\cite{1405.0412} we carried out the analysis for $N=1$ and found
12 isolated solutions to the $V=0$ equation. Using \refb{epr1} and
\refb{epr2} we see that this is in perfect agreement with the results from the
dual description. In this paper we shall carry out the computation for $N=1$, 2 and
3 using a slightly different method. The reanalysis of the $N=1$ case
is important since the sign of the ${\rm Tr} \Big( Z^{(13)} Z^{(34)}Z^{(41)}\Big)$
term in $\WW_2$ given in \refb{ew2gen}
differs from that used in \cite{1405.0412}.  It turns out however that the result
remains unchanged.

\sectiono{Supersymmetric ground states for $N=1$} \label{s2.5}

Our goal is to determine the manifold $\MM$ parametrizing the gauge 
inequivalent solutions to the equation $V=0$. Since the potential is
a sum of positive semi-definite terms, 
this in turn requires that each term in the potential vanishes separately. We shall
now analyze the condition for vanishing of various terms separately. Our method differs
from the one used in \cite{1405.0412} in that in \cite{1405.0412} we used gauge
invariant combinations of variables for analyzing the F-term equations, whereas here
we work in a specific gauge. While for $N=1$ both approaches are equally efficient,
working in a fixed gauge seems to make the analysis simpler for $N\ge 2$.

\subsection{F-term equations}

Vanishing of the F-term
contribution to the potential requires that $\p\WW/\p\vp_\alpha$ 
vanishes for each $\alpha$.
The $\p\WW/\p\Phi^{(k)}_m=0$ equations give
\be \label{eqzklgenxx}
Z^{(k\ell)}  Z^{(\ell k)} =
 -c^{(k\ell)} 
\quad \hbox{for} \quad 1\le k, \ell \le 4, \quad k\ne \ell \, .
\ee
The $\p\WW/\p Z^{(k\ell)}=0$ equations give
 \ben \label{ezeqgenxx}
&& \sum_{m=1}^3 \ve^{k\ell m} \Big( Z^{(\ell k)} \, \Phi^{(k)}_m -\Phi^{(\ell)}_m \, Z^{(\ell k)}
\Big)
+ \sum_{m=1\atop m\ne k,\ell}^4 Z^{(\ell m)} Z^{(mk)}
(-1)^{\delta_{k1} \delta_{\ell 3} \delta_{m4} } = 0  
\nonumber \\ && \qquad \qquad \qquad \qquad \qquad \qquad 
\quad \hbox{for} \quad 
1\le k, \ell \le 3, \quad k\ne \ell \, , \nonumber \\
&& \Big(\Phi^{(k)}_k  Z^{(k 4)}  - Z^{(k 4)} \Phi^{(4)}_k\Big)
+ \sum_{\ell=1\atop \ell \ne k}^3 Z^{(k\ell)} Z^{(\ell 4)} (-1)^{\delta_{k1} \delta_{\ell 3}} = 0 \quad\hbox{for} \quad
1\le k \le 3\, ,
\nonumber \\
&& \Big(Z^{(4 k)} \, \Phi^{(k)}_k    - \Phi^{(4)}_k \, Z^{(4 k)} \Big) 
+ \sum_{m=1\atop m \ne k}^3  Z^{(4m)} Z^{(mk)} (-1)^{\delta_{m1} \delta_{k 3}}= 0 \quad \hbox{for} \quad
1\le k \le 3\, .
\een

Since the superpotential has a large group of symmetries,
in order to find solutions to these F-term equations we need to choose
one representative from each symmetry orbit.
First we use the shift symmetries \refb{eflatgen} to choose
\be  \label{egauge1xx}
\Phi^{(1)}_1=0, \quad \Phi^{(1)}_2=0, \quad \Phi^{(1)}_3=0, \quad 
\Phi^{(2)}_1=0, \quad \Phi^{(2)}_2=0, \quad \Phi^{(3)}_3=0\, .
\ee 
This effectively removes the flat directions associated with the shift
symmetries.
As discussed below \refb{eflatgen}, once we have made this choice,
we can
forget about the flat directions associated with these shift symmetries and
also the fermionic superpartners of these flat directions. The information about 
$ P (y)$ is contained in the rest of the system of equations.

Next we note that the superpotential is invariant under the complexified
gauge transformation
\ben \label{ecomplexxx}
&& Z^{(k\ell)} \to a_k (a_\ell)^{-1}   \, Z^{(k\ell)}
\quad \hbox{for} \quad  1\le i\le 4, \quad 1\le k\le 4\, ,
\een
where $a_k$ for $1\le k\le 4$ are complex numbers. 
Therefore solutions to the F-term equations come as orbits of
this symmetry group. Using this we shall now make a convenient choice of gauge.
Of the four parameters encoded in
$a_1$, $a_2$, $a_3$ and $a_4$, one combination does not act 
on the fields. Therefore only three are independent. 
Now, 
since $Z^{(k\ell)}$'s for $1\le k,\ell\le 4$ are complex numbers, 
\refb{eqzklgenxx} shows that neither $Z^{(k\ell)}$ nor $Z^{(\ell k)}$ can vanish as long
as $c^{(k\ell)}$'s are chosen to be non-zero. 
This allows us to fix the gauge corresponding to the
transformations generated by $a_1$, $a_3$ and $a_4$ by setting
\be \label{egauge2xx}
Z^{(12)}=1, \quad  Z^{(23)}=1, \quad Z^{(14)}=1\, .
\ee

We now substitute the gauge choices \refb{egauge1xx} and \refb{egauge2xx} 
into the F-term equations and
look for solutions to these equations. For this we choose random rational
values of the $c^{(k\ell)}$'s for $1\le k,\ell\le 4$. A numerical
analysis of the Hilbert series using {\sc Singular}\cite{singular} and
Macaulay2\cite{mac}, 
treating the F-term equations as ideals of the ring, 
shows that the solution space is a collection of 12 points.
Explicit numerical solutions in Mathematica\cite{math} 
also yields precisely 12 solutions
for each choice of the constants $\{c^{(k\ell)}\}$.

\subsection{D-term equations}

Next we turn to solving the 
D-term equations. For this we recall that the F-term
equations do not give isolated solutions, but give orbits of the complexified
gauge group generated by the complex numbers $a_1$, $a_2$, $a_3$ and
$a_4$.
The D-term equations only respect a subgroup of this symmetry group
consisting of physical gauge transformations.
Therefore for each of the solutions to the F-term equations we can examine the
orbit under \refb{ecomplexxx} and then try to determine $a_1$, $a_2$, $a_3$
and $a_4$ by demanding that the D-term equations are satisfied. 
As before, one combination of $a_1$, $a_2$, $a_3$ and $a_4$ 
does not act on
the variables, and so we can restrict to transformations for which $a_3=1$.
However
this is not expected to fix the parameters $a_1$, $a_2$, $a_3$ and
$a_4$ completely since a subgroup
of the transformations \refb{ecomplexxx}, corresponding to physical gauge
transformations, generates a symmetry of the D-term equations of motion as
well. This subgroup is generated by
taking the $a_i$'s to be phases. This means that once we have found a set of $a_i$'s
that give a solution to the D-term
equations, we can generate other solutions by performing $U(1)^4$ 
transformations. The 
effect of these transformations will be to transform the parameters
$a_1$, $a_2$, $a_3$ and $a_4$ by
\be \label{ephasexx}
a_k \to e^{i\phi_k} a_k\, ,
\ee
where $\phi_k$ are real numbers. 
Therefore for finding solutions up to gauge transformations,
we can use the transformations \refb{ephasexx} to fix the `gauge' for 
$a_1$, $a_2$, $a_3$ and $a_4$. We choose a gauge in which
all the $a_k$'s are real and positive.  

We now transform each of the 12 solutions to the F-term equations by 
\refb{ecomplexxx} with real positive $a_k$'s 
and try to determine the $a_k$'s by solving
the D-term equations.  For each of the 12 cases, we find that there is a 
unique choice of real positive $a_k$'s
that solves the D-term equations. This shows that up to gauge transformations,
there are precisely 12 solutions to the F- and D-term equations. More accurately
we have 12 different gauge orbits, generated by $U(1)^4$
transformation, as solutions to the F- and D-term equations.

\subsection{$X^{(k)}_i$ dependent terms}

Although at the end of appendix \ref{sC} we shall 
give a general argument that all $X^{(k)}_i$'s must
vanish up to the shift symmetry described in the last line of \refb{eflatgen}, 
we shall now verify this explicitly for this example. For this
we have to demand the vanishing of the $X^{(k)}_i$ dependent terms in the 
potential given in
\refb{egaugegen}.
Since vanishing of F- and D-terms is a necessary condition for a
supersymmetric
vacuum, the choice of $Z^{(k\ell)}$'s and $\Phi^{(k)}_i$'s must be restricted to the
12 solutions that we have found. Now we note that since \refb{egaugegen}  
is a sum of 
positive definite terms, 
in order for $V_{gauge}$ to vanish, each term 
in the potential must vanish. In particular this will require
\be \label{ereqxx}
(X^{(k)}_i  - X^{(\ell)}_i) \, Z^{(k\ell)} 
= 0\, , \quad \hbox{for} \quad 1\le k,\ell\le 4, \quad k\ne \ell, 
\quad 1\le i\le 3\, .
\ee
Using the
result that the $Z^{(k\ell)}$ are non-zero, we get
$X^{(k)}_i=X^{(\ell)}_i$. Using the shift symmetry given in the third line of
\refb{eflatgen} we can set $X^{(1)}_i$ to zero for $1\le i\le 3$. As a result
$X^{(k)}_i$ for $k=2,3$ and 4 must also vanish. 
This shows that the only way to make the potential
vanish is to have all the $X^{(k)}_i$'s vanish for $1\le k\le 4$ and $1\le i\le 3$.

\subsection{Gauss' law constraint} \label{sgauss}

Finally we turn to the Gauss' law constraint. This is not a constraint on the
classical configuration but on the quantum vacuum. When we dimensionally
reduce the 3+1 dimensional theory to 0+1 dimensions we also get 
non-dynamical fields from the zeroth component of the gauge fields 
which we have set to zero. The equations of motion for $A^{(k)}_0$ imposes the
constraint that the total gauge charge carried by the state must vanish.
Since for a charged complex scalar the gauge charge will involve a product
of the field and its time derivative, the gauge charge will vanish for a
classical solution which is time independent. But we still need to examine
if the quantum ground state -- which can be regarded as the ground state
of a system of coupled harmonic oscillators describing small oscillations around
the isolated vacua, and a set of flat directions generated by the gauge
transformation -- is invariant under the gauge transformation.

Now as already mentioned, the solutions to the potential minimization equations
generate 12 different gauge orbits. The effect of gauge transformation is to move
a point in the vacuum manifold along the gauge orbit. Therefore the requirement
that the quantum ground state is gauge invariant simply translates to the
requirement that the ground state wave-function is independent of the coordinates
along the gauge orbit. 
By expressing the
kinetic term for each variable ($Z^{(k\ell)}$'s and $\Phi^{(k)}_i$'s) in terms 
of the collective coordinates generated by the gauge transformations we can
bring the Lagrangian of the collective coordinates to that of 
three independent free particles (corresponding to three independent gauge
transformations that act non-trivially on the fields) with compact target space
and positive masses.
Therefore the gauge invariant state indeed is the
lowest energy state of the system. This leads us to 
conclude that for each of the 12 gauge orbits, the quantum
ground state is invariant under gauge transformations and hence satisfies the
Gauss' law constraint.

Therefore we see from \refb{espectrum} that the  for 
$N=1$ we have
\be 
 P (y)=12\, .
\ee 
The $y$ independence of $ P (y)$ is the result of the gauge inequivalent 
solutions being isolated points, and is
consistent with the conjecture that all the
microstates carry zero angular momentum.

\sectiono{Supersymmetric ground states for $N=2$} \label{s3}

We shall now determine the number of supersymmetric ground states for $N=2$ following the
same steps as in \S\ref{s2.5}.

\subsection{F-term equations}

Vanishing of the F-term
contribution to the potential requires that $\p\WW/\p\vp_\alpha$ vanishes 
for each $\alpha$.
The $\p\WW/\p\Phi^{(k)}_m=0$ equations give
\ben \label{eqzklgen}
Z^{(k\ell)}  Z^{(\ell k)} &=&  
 -c^{(k\ell)} 
\quad \hbox{for} \quad 1\le k, \ell \le 3, \quad k\ne \ell \, ,  \nonumber \\
Z^{(k4)}Z^{(4k)} &=& - 2\, c^{(k4)}  \, ,
\quad 1\le k \le 3\, , \nonumber \\
Z^{(4k)}Z^{(k4)} &=& - c^{(k4)} \,   I_{2} - \sum_{\ell, m=1}^3 \ve^{k\ell m} \Phi^{(4)}_\ell \Phi^{(4)}_m\, ,
\quad 1\le k \le 3\, .
\een
The equations in the second line follow from the trace of the equation in the third line, but
we have listed them separately as they will be useful in analyzing the solutions.
The\break \noindent $\p\WW/\p Z^{(k\ell)}=0$ equations give
 \ben \label{ezeqgen}
&& \sum_{m=1}^3 \ve^{k\ell m} \Big( Z^{(\ell k)} \, \Phi^{(k)}_m -\Phi^{(\ell)}_m \, Z^{(\ell k)}
\Big)
+ \sum_{m=1\atop m\ne k,\ell}^4 Z^{(\ell m)} Z^{(mk)}  
(-1)^{\delta_{k1} \delta_{\ell 3} \delta_{m4} } 
= 0 \nonumber \\ && \hskip 3in \quad \hbox{for} \quad
1\le k, \ell \le 3, \quad k\ne \ell \, , \nonumber \\
&& \Big(\Phi^{(k)}_k  Z^{(k 4)}  - Z^{(k 4)} \Phi^{(4)}_k\Big)
+ \sum_{\ell=1\atop \ell \ne k}^3 Z^{(k\ell)} Z^{(\ell 4)} (-1)^{\delta_{k1} \delta_{\ell 3}  } 
= 0 \quad \hbox{for} \quad
1\le k \le 3\, ,
\nonumber \\
&& \Big(Z^{(4 k)} \, \Phi^{(k)}_k    - \Phi^{(4)}_k \, Z^{(4 k)} \Big) 
+ \sum_{m=1\atop m \ne k}^3  Z^{(4m)} Z^{(mk)} (-1)^{\delta_{m1} \delta_{k 3}  } = 0 \quad \hbox{for} \quad
1\le k \le 3\, .
\een
As in \S\ref{s2.5}, we use the shift symmetries \refb{eflatgen} to choose
\be  \label{egauge1}
\Phi^{(1)}_1=0, \quad \Phi^{(1)}_2=0, \quad \Phi^{(1)}_3=0, \quad 
\Phi^{(2)}_1=0, \quad \Phi^{(2)}_2=0, \quad \Phi^{(3)}_3=0\, .
\ee 

Next we note that the superpotential is invariant under the complexified
gauge transformation
\ben \label{ecomplex}
&& Z^{(k\ell)} \to a_k (a_\ell)^{-1}  Z^{(k\ell)}, \quad Z^{(4k)}\to (a_k)^{-1} 
M\,  Z^{(4k)}, \quad  Z^{(k4)} \to a_k \, Z^{(k4)} M^{-1}\, , \nonumber \\
&& \hskip 3.2in 
\hbox{for} \quad 1\le k\le 3, \quad k\ne \ell \, , \nonumber \\
&& \Phi^{(k)}_i\to \Phi^{(k)}_i, \quad \Phi^{(4)}_i \to M \Phi^{(4)}_i
M^{-1}\, , \quad \hbox{for} \quad  1\le i\le 3, \quad 1\le k\le 3\, ,
\een
where $a_k$ for $1\le k\le 3$ are complex numbers and $M$ is a $2\times 2$
complex matrix. Therefore solutions to the F-term equations come as orbits of
the symmetry group. Using this we shall now make a convenient choice of gauge.
Since $Z^{(k\ell)}$'s for $1\le k,\ell\le 3$ are complex numbers the first equation in
\refb{eqzklgen} shows that neither $Z^{(k\ell)}$ nor $Z^{(\ell k)}$ can vanish as long
as $c^{(k\ell)}$'s are chosen to be non-zero. 
This allows us to fix the gauge corresponding to the
transformations generated by $a_1$ and $a_3$ by setting
\be \label{egauge2}
Z^{(12)}=1, \quad  Z^{(23)}=1\, .
\ee
Similarly, since $Z^{(k4)}$ are two component
row vectors and $Z^{(4k)}$ are two component column vectors for $1\le k\le 3$,
the second equation in \refb{eqzklgen} tells us that neither $Z^{(k4)}$ nor $Z^{(4k)}$
can have both components vanishing. 
This allows us to use the transformation
generated by $M$ to set
\be \label{egauge3}
Z^{(14)} = \begin{pmatrix} 1 & 0 \end{pmatrix}\, .
\ee 
This does not fix $M$ completely since the choice of $M$ given by
\be \label{eres}
\begin{pmatrix} 1 & 0 \cr r & s \end{pmatrix}\, .
\ee
preserves the form of $Z^{(14)}$. Since $Z^{(24)}$ is non-zero, we can use this
residual gauge symmetry to set
\be \label{egauge4}
Z^{(24)} = \begin{pmatrix} 0 & 1  \end{pmatrix}\, .
\ee
There is one case where this gauge condition fails, and that is in
the case when $Z^{(24)}$ is parallel to $Z^{(14)}$ to begin with. In this case
the gauge symmetry described in \refb{eres} cannot be used to bring
$Z^{(24)}$ to the form \refb{egauge4}. We shall deal with this case
separately. Once we have used $a_1$, $a_3$ and $M$ to fix these gauges,
we cannot use $a_2$ any more since its action is determined in terms
of the others.

We now substitute the gauge choices \refb{egauge1}, \refb{egauge2}, 
\refb{egauge3} and \refb{egauge4} into the F-term equations and
look for solutions to these equations. For this we choose random rational
values of the $c^{(k\ell)}$'s for $1\le k<\ell\le 4$. A numerical
analysis of the Hilbert series using {\sc Singular} and Macaulay2, 
treating the F-term equations as ideals of the ring, 
shows that the solution space is a collection of 56 points.
Explicit numerical solutions in Mathematica also yields precisely 56 solutions
for each choice of the constants $\{c^{(k\ell)}\}$.

We also explore the possibility of having solutions with $Z^{(24)}$ proportional
to $Z^{(14)}$ for which the gauge choice \refb{egauge4} will be invalid. In this case
the analysis of the Hilbert series shows that there are no solutions. Explicit attempts
to find solutions to the F-term equations in Mathematica also gives no results.
This shows that there are no solutions for which the gauge choice
\refb{egauge4} breaks down.

\subsection{D-term equations}

The solutions to the F-term
equations give orbits of the complexified
gauge group generated by the complex numbers $a_1$, $a_2$, $a_3$ and
the $2\times 2$ complex matrix $M$ as given in \refb{ecomplex}.
The D-term equations only respect a subgroup of this symmetry group
consisting of physical gauge transformations.
Therefore for each of the solutions to the F-term equations we can examine the
orbit under \refb{ecomplex} and then try to determine $a_1$, $a_2$, $a_3$
and $M$ by demanding that the D-term equations are satisfied. 
As before, one combination of $a_1$, $a_2$, $a_3$ and $\det M$ 
do not act on
the variables, and so we can restrict to transformations for which $\det M=1$.
However
this is not expected to fix the parameters $a_1$, $a_2$, $a_3$ and
$M$ completely since a subgroup
of the transformations \refb{ecomplex}, corresponding to physical gauge
transformations, generates a symmetry of the D-term equations of motion as
well. This subgroup is generated by
taking the $a_i$ to be phases and $M$ to be an
$SU(2)$ matrix. This means that once we have found a set of $a_i$'s and $M$ 
that give a solution to the D-term
equations, we can generate other solutions by performing $U(1)^3\times 
SU(2)$ transformations. The 
effect of these transformations will be to transform the parameters
$a_1$, $a_2$, $a_3$ and $M$ by
\be \label{ephase}
a_k \to e^{i\phi_k} a_k, \quad M \to U \, M\, ,
\ee
where the $\phi_k$ are real numbers and $U$ is an SU(2) matrix. 
Therefore for finding solutions up to gauge transformations,
we can use the transformations \refb{ephase} to fix the `gauge' for 
$a_1$, $a_2$, $a_3$ and $M$. Using the transformations generated by the
$\phi_k$'s we can make the $a_k$'s real and positive.  Furthermore, one can easily
check that using the transformations generated by $U$ and the constraint
$\det M=1$, we can bring $M$ to the form
\be \label{eformm}
M = \begin{pmatrix} 1/a & b \cr 0 & a \end{pmatrix}\, ,
\ee
where $a$ is a real positive number and $b$ is a complex number.

We now transform each of the 56 solutions to the F-term equations by 
\refb{ecomplex} with real positive $a_k$'s and $M$ of the form
\refb{eformm} and try to determine the $a_k$'s, $a$ and $b$ by 
numerically solving
the D-term equations.  For each of the 56 cases, we find that there is a 
unique choice of real positive $a_k$'s and $M$ of the form \refb{eformm}
that solves the D-term equations. This shows that up to gauge transformations,
there are precisely 56 solutions to the F- and D-term equations. More accurately
we have 56 different gauge orbits, generated by U(1)$^3\times$SU(2)
transformation, as solutions to the F- and D-term equations.

\subsection{$X^{(k)}_i$ dependent terms}

Next we have to check if the $X^{(k)}_i$ dependent terms in the 
potential given in
\refb{egaugegen} vanish.
For each of the 56 solutions there is a simple way to make these terms
vanish -- we simply choose $X^{(k)}_i=0$ for $1\le i\le 3$, $1\le k\le 4$.
The question we shall be interested in is: are there other configurations
that make the $X^{(k)}_i$ dependent terms vanish?

Since vanishing of F- and D-terms is a necessary condition for a 
supersymmetric
vacuum, the choice of $Z^{(k\ell)}$'s and $\Phi^{(k)}_i$'s must be restricted to the
56 solutions that we have found. We shall now argue that for each of
these solutions, the
$X^{(k)}_i$'s for $1\le k\le 4$ and $1\le i\le 3$ 
must vanish identically
up to the shift symmetries described in the
last line of \refb{eflatgen}. For this we turn to the 
potential \refb{egaugegen} and note that since this is a sum of positive definite terms, 
in order for $V_{gauge}$ to vanish, each term 
in the potential must vanish. In particular this will require
\be \label{ereq}
X^{(k)}_i  Z^{(k\ell)} - Z^{( k\ell)} 
X^{(\ell)}_i = 0\, , \quad \hbox{for} \quad 1\le k,\ell\le 4, \quad k\ne \ell, 
\quad 1\le i\le 3\, .
\ee
For $1\le k,\ell\le 3$, $k\ne \ell$, 
$X^{(k)}_i$ and   $Z^{(k\ell)}$ are numbers and hence, using the
result that the $Z^{(k\ell)}$ are non-zero, we get
$X^{(k)}_i=X^{(\ell)}_i$. Using the shift symmetry given in the third line of
\refb{eflatgen} we can set $X^{(1)}_i$ to zero for $1\le i\le 3$. As a result
$X^{(2)}_i$ and $X^{(3)}_i$ must also vanish. Choosing $k=4$ and $\ell=1$, 2
or 3  
in \refb{ereq}, and vice versa,  we now get
\be \label{ereq1}
X^{(4)}_i Z^{(4\ell)} = 0, \quad Z^{(\ell 4)} X^{(4)}_i = 0\, .
\ee
Since we have seen that $Z^{(4\ell)}$ and $Z^{(\ell 4)}$ cannot have their both components 
vanish, this shows that $Z^{(4\ell)}$ and $Z^{(\ell 4)}$ are right and left
eigenvectors of $X^{(4)}_i$ with zero eigenvalues.  Let us suppose that $X^{(4)}_i$ is
non-zero for at least one $i$. 
In that case the $Z^{(\ell 4)}$'s for $1\le \ell \le 3$
must be proportional to each other since each of them
is a left eigenvector of the non-vanishing $X^{(4)}_i$ with zero eigenvalue.
However as already mentioned,
while examining the validity of the gauge choice \refb{egauge4} we have
explicitly checked that there are no solutions to the F-term equations in 
which $Z^{(14)}$ and $Z^{(24)}$ are parallel to each other. This shows that
our initial assumption must have been wrong, and $X^{(4)}_i$ for each $i$
must vanish identically. This shows that the only way to make the potential
vanish is to have all the $X^{(k)}_i$'s vanish for $1\le k\le 4$ and $1\le i\le 3$.

\subsection{Gauss' law constraint}

Finally we turn to the Gauss' law constraint. This analysis is identical to that given in
\S\ref{sgauss} except for one difference: the collective modes associated with the
SU(2) gauge transformations have a kinetic term given by that of a rigid rotator with
positive definite inertia matrix instead of that of a free particle with positive mass.
Positive definiteness of the inertia matrix guarantees 
that the ground state wave-function is independent of these 
collective coordinates and hence the ground state is gauge invariant. Thus the Gauss'
law constraint is automatically satisfied for each of the 56 gauge orbits.

Therefore we see from \refb{espectrum} that for 
$N=2$ we have
\be 
 P (y)=56\, .
\ee 
As in \S\ref{s2.5},
the $y$ independence of $ P (y)$ is the result of the gauge inequivalent 
solutions being isolated points, and is
consistent with the conjecture that all the
microstates carry zero angular momentum.
We also see from \refb{epr1} and \refb{epr2} that the 
counting in the dual description gives  a BPS index 56. 
Thus there is perfect agreement between our result and that in the dual
description.

\sectiono{Supersymmetric ground states for $N=3$} \label{s4}

The analysis of the equations for the $N=3$ case proceeds similarly to that for the $N=2$ case.
Up to \refb{egauge2} there is essentially no change. The gauge conditions \refb{egauge3} and
\refb{egauge4} are replaced by
\be  \label{enewg}
Z^{(14)}=(1,0,0), \quad Z^{(24)}=(0,1,0), \quad Z^{(34)}=(0,0,1)\, .
\ee
Such a gauge choice is always possible if initially the 
vectors $Z^{(14)}$, $Z^{(24)}$ and $Z^{(34)}$ are linearly independent.
The case where they are linearly dependent is analyzed separately and we find no
solution in this sector. With the gauge choice \refb{enewg} we find 208 distinct solutions
to the set of F-term constraints. Furthermore for each of these solutions the $X^{(k)}_i$'s
can be shown to vanish identically using arguments identical to those given
below \refb{ereq1}. 
We have not checked explicitly that 
for each of these solutions we have a unique solution to the D-term constraints
up to gauge transformation, but since the D-term constraints usually amount
to quotienting by complexified gauge transformations, and since following arguments
similar to the one given below \refb{egauge2} one can argue that 
none of the vectors $Z^{(4k)}$ and $Z^{(k4)}$ can vanish as a vector, one
expects on general grounds that the quotient by complexified gauge transformation
will give a unique solution for each solution to the F-term constraints.
Therefore we conclude that in this case we have
\be \label{e208}
 P (y)=208\, .
\ee
Again the $y$ independence of $ P (y)$ is the result of the gauge 
inequivalent 
solutions being isolated points, and is
consistent with the conjecture that all the
microstates carry zero angular momentum. \refb{e208} is
in perfect agreement with  
the result in the dual description which, according to \refb{epr1}
and  \refb{epr2},
gives $ P (-1)=208$.

\sectiono{Conclusion} \label{sconc}

In this paper we have provided evidence for the conjecture that all the microstates
of single centered BPS black holes carry strictly zero momentum at a generic point in 
the moduli space. This conjecture
is consistent with the near horizon $AdS_2\times S^2$ geometry of extremal black
holes, but there are no direct arguments in the microscopic theory leading to this 
conjecture. Therefore the test of this conjecture provides evidence that the black hole
horizon carries more information than just some average properties of the microstates. 

These results put a strong
constraint on possible fuzzball solutions describing black hole 
microstates\cite{0502050,0810.4525,0701216,0804.0552,0811.0263}.
Typical fuzzball
solutions are constructed at special points in the moduli space and carry 
states of different angular momenta.  
For demonstrating that they describe genuine black
hole microstates, one needs to construct these solutions for generic values of
the asymptotic moduli and show that the solutions so obtained carry strictly
zero angular momentum.

A different approach to this problem has been suggested in 
\cite{1507.06670}
where one constructs 
solutions with asymptotic $AdS_2$ boundary conditions. These solutions 
carry
strictly zero angular momentum.
However these are not truly in the spirit of the fuzzball program since 
they exist within the near horizon geometry of the black hole instead of replacing
the near horizon geometry by a smooth solution. Furthermore since these
solutions have two asymptotic boundaries, they most likely describe an entangled
state living on two copies of the black hole Hilbert space instead of the
microstates of a single black hole\cite{1101.4254}.

\bigskip

\noindent {\bf Acknowledgements:} 
We wish to thank Atish Dabholkar, Joao Gomes, Dileep Jatkar, Jan Manschot, 
Noppadol Mekareeya,
Sameer Murthy,
Boris Pioline and Savdeep Sethi for useful discussions.
We have benefited from the use of symbolic manipulation programs
Macaulay2, Mathematica and {\sc Singular}, and the use of the
High Performance Cluster 
Computing facility at the Harish-Chandra Research Institute.
This work was
supported in part by the 
DAE project 12-R\&D-HRI-5.02-0303. 
The work of A.S. was also supported in
part by the
J. C. Bose fellowship of 
the Department of Science and Technology, India.

\appendix

\sectiono{Normalization of $Z$-$Z$-$Z$ coupling} \label{sC}

In this appendix we shall determine the normalizations and signs of the
$Z$-$Z$-$Z$ coupling appearing in \refb{ew2gen} by analyzing respectively
open
string amplitudes and symmetry requirements. 
The computation will be done around a background 
in which all the circles of $T^6$ are orthonormal and have radius 
$\sqrt{\alpha'}$. Fluctuations away from this background will be
parametrized by the constants $c^{(k\ell)}$ and $c^{(k)}$ appearing in
\refb{ew3gen} and \refb{evdgen}. Since we work in the region where these
constants are small (see footnote \ref{f77}), 
the corrections to the cubic terms in the superpotential
proportional 
to these constants can be ignored.

The easiest way to determine a cubic term in the superpotential is to examine
the Yukawa coupling between two fermions and one boson that arises from
this term. For this we 
need to construct the vertex operators of the corresponding states and 
compute their three point function on the disk. We shall denote by $b$ and $c$
the usual diffeomorphism ghost fields, by $\beta$ and $\gamma$ the superconformal
ghosts and by $\phi$ the scalar that arises from bosonization of the
$\beta-\gamma$ system\cite{FMS}, normalized such that
\be \label{eghost}
\langle c(z_1) e^{-\phi}(z_1) \, c(z_2)e^{-\phi/2}(z_2)  \, c(z_3)e^{-\phi/2}(z_3)\rangle
= (z_1-z_2)^{1/2} (z_1-z_3)^{1/2}  (z_2-z_3)^{3/4} \, ,
\ee 
up to a sign. In the matter sector,
we shall combine the compact spatial coordinates
into complex coordinates as
\be 
w^1 = x^4 + i x^5, \quad w^2 = x^6+i x^7, \quad w^3 = x^8+ix^9\, .
\ee
$w^1,\ldots,w^3$ are complex coordinates.
For each coordinate $w^i$ we have a complex world-sheet scalar
field which we shall denote by $W^i$. Their superpartners are complex
world sheet fermions which we denote by $\psi^i$. We also
introduce the complex spin field $s_i$ that twists $\psi^i$ by a 
$Z_2$ transformation $\psi^i\to - \psi^i$ 
and the real twist field $\sigma_i$ that twists 
$W^i$ by a $Z_2$ transformation $W^i\to - W^i$. Since a world-sheet scalar
satisfying a Neumann boundary condition at one end and a Dirichlet boundary
condition at the other end has a half-integer mode expansion,
the twist fields $\sigma_i$ 
will be necessary for constructing the vertex operators for open string states
satisfying Neumann-Dirichlet boundary conditions in some directions.
The spin fields $s_i$ will be needed for constructing the vertex 
operators for open string states
satisfying Neumann-Dirichlet boundary conditions in some directions and also
for constructing vertex operators in the Ramond sector. 
Finally we shall denote by $s^{(nc)}_\alpha$ the 
spin fields
associated with the non-compact directions carrying spinor index $\alpha$
of $SO(3,1)$.
We shall use standard
normalizations for the fields $\psi^i$, $s_i$ and $\sigma_i$, e.g.
\ben \label{eope}
&& \psi^i(z_1) \bar \psi^j(z_2) = \delta_{ij} \, (z_1-z_2)^{-1}, \quad s_i(z_1)
\bar s_j(z_2) = \delta_{ij} \, (z_1-z_2)^{-1/4}, 
\nonumber \\
&& \bar \psi^i (z_1) s_j (z_2) = \delta_{ij} \, (z_1-z_2)^{-1/2} 
\bar s_j(z_2), \quad
\psi^i (z_1) \bar s_j (z_2) =  \delta_{ij} \, (z_1-z_2)^{-1/2} s_j (z_2), 
\nonumber \\ &&
\sigma_i(z_1) \sigma_j(z_2) = \delta_{ij} \, (z_1-z_2)^{-1/4}, 
\quad s^{(nc)}_\alpha (z_1) s^{(nc)}_\beta (z_2) = \eps_{\alpha\beta}
(z_1-z_2)^{-1/2}\, ,
\een
up to multiplicative signs and less singular additive terms. The operator
products given in the last line are determined by the conformal weights
of $\sigma_i$, $s^{(nc)}_\alpha$ and their normalizations. The operator 
products in the first two lines can be shown to be mutually compatible by 
bosonizing the fermions $\psi^i$ to scalar fields $\phi_i$ and using the
identification
\be
\psi^i = e^{i\phi_i}, \quad \bar\psi^i = e^{-i\phi_i}, \quad s_i =
e^{i\phi_i/2}, \quad \bar s_i = e^{-i\phi_i/2}\, .
\ee

The other ingredient we need for the construction of the vertex operator is
Chan-Paton factors. Since we have altogether $N+3$ D-branes, the
Chan-Paton factors can be taken to be $(N+3)\times (N+3)$ matrices. We
shall choose the convention in which the first three rows and columns
represent the branes 1, 2 and 3 and the last $N$ rows and columns 
label the brane stack 4. In this notation, for $1\le k,\ell\le 3$, $1\le r,s\le N$,
the Chan-Paton factors for the 
  vertex operators for $\Phi^{(k)}_\ell$, $(\Phi^{(4)}_\ell)_{rs}$, $Z^{(k\ell)}$,
  $Z^{(k4)}_r$ and $Z^{(4k)}_r$ will be matrices whose only
non-zero entries are the $(k,k)$'th, $(3+r,3+s)$'th, $(k,\ell)$'th, 
$(k,3+r)$'th and $(3+r,k)$'th elements, respectively.
In the analysis that follows
we shall
suppress the Chan-Paton factors -- they just accompany the vertex operators
as multiplicative matrices. The correlation function of a given set of vertex
operators will contain a term proportional to 
the trace of the ordered product of their
Chan-Paton factors.

The vertex operators for the states corresponding to $\Phi^{(k)}_i$ and
$Z^{(k\ell)}$ can be constructed using these operators. For a superfield
$A$ we shall denote by $V^f_{A,\alpha}$ and $V^b_A$ the vertex operators of
the fermionic and bosonic components of the superfield. Here $\alpha$
denotes a 3+1 dimensional
spinor index. In this notation we have,
for example
\ben \label{ea.4}
&& V^b_{\Phi^{(k)}_1} = c \, e^{-\phi} \psi^1, \quad V^f_{\Phi^{(k)}_1,\alpha} 
= c \, e^{-\phi/2} s_1 \bar s_2 \bar s_3 s^{(nc)}_\alpha,
\nonumber \\
&& V^b_{Z^{(12)}} = c \, e^{-\phi} \sigma_1 \sigma_2 
s_1 s_2, \quad V^f_{Z^{(12)},\alpha}
= c\, e^{-\phi/2} \sigma_1  \sigma_2 \bar s_3 s^{(nc)}_\alpha\, ,
\een
up to multiplicative signs. Here the $\psi^1$ factor in the expression for
$V^b_{\Phi^{(k)}_1}$ reflects that this mode represents position / Wilson
line on the brane along $w^1$, whereas the $s_1 \bar s_2 \bar s_3$ factor
in
$V^f_{\Phi^{(k)}_1,\alpha}$ is obtained by starting with the operator 
$\bar s_1 \bar s_2 \bar s_3$ that appears in the expression for the
supersymmetry generator, and taking the leading term in its operator
product with $\psi^1$ -- the same factor that is present in $V^b_{\Phi^{(k)}_1}$.
In the expression for $V^b_{Z^{(12)}}$ the $\sigma_1 \sigma_2$ and 
$s_1  s_2$
factors reflect that the fields $W^1$ and $W^2$ and their fermionic partners
$\psi^1$ and $\psi^2$ satisfy Neumann boundary conditions at one end of the
open string and Dirichlet boundary conditions at the other end. On the other hand
the $\sigma_1  \sigma_2 \bar s_3$ term in the expression for $V^f_{Z^{(12)},\alpha}$
is the leading term in the expression for the operator product of 
$\bar s_1 \bar s_2 \bar s_3$ and $\sigma_1 \sigma_2 s_1 s_2$.  
Following this strategy we can find the expressions for all the vertex operators
$V^b_{\Phi^{(k)}_i}$, $V^f_{\Phi^{(k)}_i,\alpha}$, $V^b_{Z^{(k\ell)}}$ and
$V^f_{Z^{(k\ell)},\alpha}$. 

Let us now calculate the three point function 
\be
\langle V^b_{\Phi^{(4)}_1} V^f_{\Phi^{(4)}_2,\alpha} V^f_{\Phi^{(4)}_3,\beta}\rangle\, ,
\ee
to fix the normalization of the terms in $\WW_4$ in \refb{ew4gen}. 
Using the generalization of \refb{ea.4} we see that this is computed 
using the correlation function
\be 
\langle c\, e^{-\phi} \psi^1 (z_1) \, c\, e^{-\phi/2} \bar s_1 s_2 \bar s_3
s^{(nc)}_\alpha  (z_2)
\, c\, e^{-\phi/2} \bar s_1 \bar s_2 s_3 s^{(nc)}_\beta (z_3)\rangle\, .
\ee
This correlator can be factorized and evaluated using \refb{eghost} and
\refb{eope} as
\ben
&& \langle c\, e^{-\phi} (z_1) \, c\, e^{-\phi/2} (z_2) \, c\, e^{-\phi/2}(z_3) \rangle_{ghost}
\, \langle \psi^1 (z_1)  \bar s_1(z_2) \bar s_1 (z_3) \rangle_{\psi^1}
 \nonumber \\ && \langle s_2 (z_2) \bar s_2 (z_3) \rangle_{\psi^2}
\langle  \bar s_3 (z_2)
s_3 (z_3)\rangle_{\psi^3} \langle
s^{(nc)}_\alpha  (z_2)
s^{(nc)}_\beta (z_3)\rangle_{nc}\nonumber \\ &=&
\eps_{\alpha\beta}\, ,
\een
up to a sign. This agrees with the normalization of the three point coupling of
$\Phi^{(4)}$ given in \refb{ew4gen} after ignoring the overall $\sqrt 2$ factor.
(Since the $\sqrt 2$ factor appears universally in front of all terms in 
the superpotential $\WW$,
we can ignore this for fixing the relative normalization between different
terms.)

Next we compute the $\Phi$-$Z$-$Z$ three point
function. For example the $\Phi^{(1)}_3 Z^{(12)} Z^{(21)}$ coupling
will be given by the coefficient of $\eps_{\alpha\beta}$ in
\ben
\langle V^b_{\Phi^{(1)}_3}(z_1) V^f_{Z^{(12)},\alpha} (z_2)
V^f_{Z^{(21)},\beta}(z_3)\rangle &=& 
\langle c\, e^{-\phi} \psi^3 (z_1) \, c\, e^{-\phi/2} \sigma_1  
\sigma_2 \bar s_3 s^{(nc)}_\alpha(z_2) 
\, c\, e^{-\phi/2} \sigma_1  \sigma_2 \bar s_3 s^{(nc)}_\beta(z_3)
\rangle\nonumber \\
&=& \eps_{\alpha\beta}
\een
up to a sign. In the last step we have used \refb{eghost} and
\refb{eope} and factorized the correlator into contributions from the ghost
sector and different components of the matter sector. 
This is in agreement with the normalization of the 
$\Phi$-$Z$-$Z$ coupling given in \refb{ew1gen}.

Finally let us compute the $Z$-$Z$-$Z$ three point coupling. For definiteness
we focus on the $Z^{(12)}
Z^{(23)} Z^{(31)}$ coupling. For this we compute
\ben
&& \langle V^b_{Z^{(12)}}(z_1) V^f_{Z^{(23)},\alpha} (z_2)
V^f_{Z^{(31)},\beta}(z_3)\rangle \nonumber \\ &=& 
\langle c\, e^{-\phi} \sigma_1 \sigma_2 s_1 s_2 (z_1) \, c\, e^{-\phi/2} \sigma_2 
\sigma_3 \bar s_1 s^{(nc)}_\alpha(z_2) 
\, c\, e^{-\phi/2} \sigma_3  \sigma_1 \bar s_2 s^{(nc)}_\beta(z_3)
\rangle\nonumber \\
&=& \eps_{\alpha\beta}
\een
Again in the last step we have used \refb{eghost} and
\refb{eope} and factorized the correlator into contributions from the ghost
sector and different components of the matter sector. 
The result is valid only up to a sign. This agrees with the
normalization of the $Z$-$Z$-$Z$ coupling in $\WW_2$ given in \refb{ew2gen}.

Following the same procedure we can show that the coefficients of all
the $\Phi$-$\Phi$-$\Phi$, $\Phi$-$Z$-$Z$ and $Z$-$Z$-$Z$ couplings are
unity up to signs. In principle these  signs can be determined by careful string 
theory
computation but we shall describe an alternative approach based on symmetry
considerations. For this we consider a more general system than what has been
discussed so far, containing 
$N_1$ D2-branes along 4-5
directions, $N_2$ D2-branes along 6-7 directions, $N_3$ D2-branes along
8-9 directions and $N_4$ D6-branes along 4-5-6-7-8-9 directions. 
We claim that the correct form of the superpotential, 
up to field redefinition,
is given by
\ben \label{ew1genxx}
\WW_1 &=&  
\sqrt 2 \, {\rm Tr} \, 
\bigg[ \left( \Phi^{(1)}_3 Z^{(12)} Z^{(21)} - \Phi^{(2)}_3 Z^{(21)} Z^{(12)}\right)
+ 
\left(\Phi^{(2)}_1 Z^{(23)} Z^{(32)} 
- \Phi^{(3)}_1 Z^{(32)} Z^{(23)}\right) \nonumber \\ && 
+ \left(\Phi^{(3)}_2  Z^{(31)} Z^{(13)}
- \Phi^{(1)}_2 Z^{(13)} Z^{(31)}\right) 
+  \left(\Phi^{(1)}_1 Z^{(14)} Z^{(41)} 
- \Phi^{(4)}_1 Z^{(41)} Z^{(14)}\right)  \nonumber \\ &&
+ \left(\Phi^{(2)}_2  Z^{(24)}  Z^{(42)} 
- \Phi^{(4)}_2 Z^{(42)} Z^{(24)}\right) +
\left(\Phi^{(3)}_3 Z^{(34)}  Z^{(43)} - \Phi^{(4)}_3 Z^{(43)} Z^{(34)}\right)
\bigg]
\, , \nonumber \\
\een
\ben \label{ew2genxx}
\WW_2 &=&   \sqrt 2\,  \Tr\bigg[
Z^{(31)} Z^{(12)}Z^{(23)} + Z^{(13)} Z^{(32)}Z^{(21)}  +  Z^{(12)} Z^{(24)}Z^{(41)}
+  Z^{(42)} Z^{(21)}Z^{(14)}  \nonumber \\ &&
-  Z^{(13)} Z^{(34)}Z^{(41)}
+  Z^{(31)} Z^{(14)}  Z^{(43)}+  Z^{(34)} Z^{(42)}Z^{(23)} 
 +  Z^{(43)} Z^{(32)}Z^{(24)} 
 \bigg]
\, ,
\een
\ben \label{ew3genxx}
\WW_3 \, =\,   \sqrt 2\, 
\Tr \bigg[ && \hskip -20pt c^{(12)} \left(\Phi^{(1)}_3 \otimes I_{N_2} 
- I_{N_1} \otimes \Phi^{(2)}_3\right) + c^{(23)} \left(\Phi^{(2)}_1 \otimes I_{N_3}- I_{N_2}
 \otimes  \Phi^{(3)}_1\right) \nonumber \\ && \hskip -20pt
+ c^{(13)} \left(\Phi^{(3)}_2 \otimes I_{N_1} - I_{N_3}  \otimes \Phi^{(1)}_2\right) 
+ c^{(14)} \left(\Phi^{(1)}_1  \otimes I_{N_4}  - I_{N_1}  \otimes\Phi^{(4)}_1\right) 
\nonumber \\ && \hskip -20pt
+ c^{(24)}\left(\Phi^{(2)}_2  \otimes I_{N_4} - I_{N_2}  \otimes \Phi^{(4)}_2\right) 
+ c^{(34)} \left(\Phi^{(3)}_3 \otimes I_{N_4} 
- I_{N_3} \otimes \Phi^{(4)}_3\right) 
\bigg]
\, , \nonumber \\
\een
and
\ben \label{ew4genxx}
\WW_4 &=& - \sqrt 2 \, \bigg[\Tr \Big( \Phi^{(1)}_1\Phi^{(1)}_2\Phi^{(1)}_3 - \Phi^{(1)}_1\Phi^{(1)}_3\Phi^{(1)}_2\Big) - \Tr \Big( 
\Phi^{(2)}_1\Phi^{(2)}_2\Phi^{(2)}_3 - \Phi^{(2)}_1\Phi^{(2)}_3\Phi^{(2)}_2\Big) 
\nonumber \\ && +
\Tr \Big(  \Phi^{(3)}_1\Phi^{(3)}_2\Phi^{(3)}_3 - \Phi^{(3)}_1\Phi^{(3)}_3\Phi^{(3)}_2\Big) 
+
\Tr \Big( \Phi^{(4)}_1\Phi^{(4)}_2\Phi^{(4)}_3 - \Phi^{(4)}_1\Phi^{(4)}_3\Phi^{(4)}_2\Big) \bigg]\, .
\een
This superpotential reduces to the one given in \refb{ew1gen}-\refb{ew4gen}
for $N_1=N_2=N_3=1$.
We shall now describe the arguments 
leading to \refb{ew1genxx}-\refb{ew4genxx}.

Let us begin with the arguments leading to the form of $\WW_2$. 
Since by field redefinitions involving changes of signs of the
$Z^{(k\ell)}$'s we can change the relative signs of various terms
in $\WW_2$, we only have to show that $\WW_2$ is given by
\refb{ew2genxx} up to these field redefinitions. Now 
these field redefinitions can only change the signs of an
even number of terms in $\WW_2$ 
since each $Z^{(k\ell)}$ appears as a factor in two of the terms.  Thus
an expression for $\WW_2$ with an even number of minus signs cannot be
turned into an expression with an odd number of minus signs and vice versa.
Furthermore, it can be shown by inspection that  by these field redefinitions
all possible choices 
of $\WW_2$ with an even number of minus signs can be brought to the form
in which each term in $\WW_2$ has positive sign, and
all possible choices of $\WW_2$ with an odd number of minus signs can be 
brought to the form given in \refb{ew2genxx}. Thus the possible candidates
for $\WW_2$ can be restricted to either \refb{ew2genxx} or the one with
all positive signs.
We shall argue shortly that requiring symmetry under the exchange of
different stacks of D-branes leads to the form given in \refb{ew2genxx}.

Next we turn to $\WW_1$ and $\WW_3$.
The
shift symmetries \refb{eflatgen} fix the relative sign between the pair of terms
inside each parenthesis in \refb{ew1genxx} and \refb{ew3genxx}, but do not
fix the signs that appear in front of the parentheses.  However
starting with any arbitrary choice of these signs, we can arrive at \refb{ew1genxx}
and
\refb{ew3genxx} by redefinition involving changes of the signs
of the fields $\Phi^{(k)}_i$ and the
parameters $c^{(k\ell)}$.  These field redefinitions change the signs
of various terms in $\WW_4$, but leave $\WW_2$ unchanged.

Finally turning to $\WW_4$ we see that the
relative sign between the pair of terms inside each parenthesis is fixed by
the requirement that these come from the dimensional reduction of
$\NN=4$ supersymmetric theories in 3+1 dimensions. 
We shall see shortly that the relative signs between the different 
parentheses are  fixed by the symmetry under the exchange of brane stacks.
This however does not fix the overall sign of $\WW_4$
leaving behind a 2-fold ambiguity. 
We expect that a careful string theory
calculation will be able to resolve this ambiguity, but we have not done this.
As we have described in the text,
the choice of $\WW_4$ given in \refb{ew4genxx},
after restriction to the case $N_1=N_2=N_3=1$, $N_4=N$ 
gives the results 12, 56 and 208 for the
index for $N=1$, 2 and 3, respectively,
in agreement with the results in the dual description. In contrast 
the opposite choice of sign gives the results 12, 60 and 232 for the index
for the cases $N=1$, 2 and 3, respectively. 
These do not agree with the results computed
using the dual description.

What remains is to show how the exchange symmetry constrains 
the form of
$\WW_2$ and $\WW_4$.
For this we need to examine how the exchange symmetry acts 
on the coefficients $c^{(k\ell)}$.
It was shown in \cite{1405.0412} that 
the coefficients $c^{(k\ell)}=c^{(\ell k)}$ for $1\le k<\ell\le 4$ are 
determined  in terms of the
background values of the metric and 2-form fields. For general
values of $N_1$, $N_2$, $N_3$ and $N_4$ the results are
as follows:\footnote{In \cite{1405.0412} only the values of $|c^{(k\ell)}|$
were determined. 
Here we have  chosen an extra minus sign in the expressions for $c^{(k\ell)}_I$ for $1\le k<\ell\le 3$
in order to have simple realization of different symmetries.}
\ben \label{esol2}
& g_{47} +  g_{56} = N_{12}\, c^{(12)}_R, \quad & g_{57}- g_{46} =   
N_{12}\, c^{(12)}_I, 
\quad g_{49} +  g_{58} =  N_{13}\, c^{(13)}_R, \quad  g _{59} - g_{48}   = 
N_{13}\, c^{(13)}_I, \nonumber \\
&   b_{68} - b_{79} = N_{14}\,  c^{(14)}_R, \quad &b_{69} + b_{78} =  N_{14}
\, c^{(14)}_I, \quad 
g_{69} +  g_{78} =  N_{23}\, c^{(23)}_R, \quad  g _{79}  - g_{68} =  N_{23}
\, c^{(23)}_I, \nonumber \\
& b_{48}  - b_{59}  =  N_{24}\, c^{(24)}_R, \quad & b_{49} + b_{58} = N_{24}
\,  c^{(24)}_I, \quad
b_{46} -  b_{57}   =  N_{34}\, c^{(34)}_R, \quad b_{47} + b_{56} =  N_{34}\, c^{(34)}_I\, ,
\nonumber \\
\een
where $N_{ij}=2(N_i+N_j)$
and
the subscripts $R$ and $I$ stand for real and imaginary parts respectively.
\refb{esol2} generalizes the result of \cite{1405.0412} for $N_1=N_2=N_3=N_4=1$
following the same logic.
For completeness we also give the expressions for the Fayet-Iliopoulos parameters
$c^{(k)}$ for $1\le k\le 4$ in terms of the background fields:
\ben \label{esoldir}
&& c^{(1)} = {1\over2} \left(b_{45}-b_{67}-b_{89}\right) + c_0, 
\quad c^{(2)} ={1\over 2} \left(
b_{67} - b_{45} - b_{89}\right) + c_0, \nonumber \\
&& c^{(3)} = {1\over 2} \left(b_{89}- b_{45} - b_{67}\right)+ c_0, 
\quad
c^{(4)} = {1\over 2} \left(b_{45} + b_{67} + b_{89}\right)+ c_0  \, ,
\een
where $c_0$ is a constant that is chosen to ensure that
$\sum_{k=1}^4 c^{(k)} N_k=0$.

Now type IIA string theory on $T^6$ has an exchange symmetry
$x^4\leftrightarrow x^6$, $x^5\leftrightarrow x^7$, 
$N_1\leftrightarrow N_2$ under which the D2-brane stacks 1 and 2
get exchanged and the stacks 3 and 4 remain unchanged. 
We see from \refb{esol2} that under
this transformation $c^{(34)}$ changes sign, $c^{(12)}$ remains unchanged, 
and $c^{(1i)}$ and $c^{(2i)}$ get exchanged for
$i=3,4$. Thus there must be an action on the variables $\Phi^{(k)}_i$ and
$Z^{(k\ell)}$ which, together with these transformations on the $c^{(k\ell)}$'s and $N_i$'s,
transform the
$\WW_i$'s at most by an overall multiplicative phase.\footnote{Multiplying 
the superpotential by an overall phase 
leaves the potential invariant.}
It is easy to verify that for the superpotential given in
\refb{ew1genxx}-\refb{ew4genxx}
the following accompanying
transformation takes $\WW_i\to -\WW_i$ for $1\le i\le 4$:
\ben \label{etr11}
&& (\Phi^{(4)}_3 , \Phi^{(3)}_3) \to  (\Phi^{(4)}_3 , \Phi^{(3)}_3), \quad 
(\Phi^{(2)}_3 , \Phi^{(1)}_3) \to (\Phi^{(1)}_3,\Phi^{(2)}_3 ), \nonumber \\ && 
(\Phi^{(2)}_1 , \Phi^{(3)}_1) \leftrightarrow (\Phi^{(1)}_2, \Phi^{(3)}_2), \quad
(\Phi^{(4)}_1 , \Phi^{(1)}_1) \leftrightarrow -(\Phi^{(4)}_2, \Phi^{(2)}_2), \nonumber \\
&& Z^{(34)}\to Z^{(34)}, \quad Z^{(i1)} \leftrightarrow - Z^{(i2)},
\quad Z^{(1i)} \leftrightarrow - Z^{(2i)}, 
\quad \hbox{for $i=3,4$}, \nonumber \\ && 
Z^{(12)}\leftrightarrow -Z^{(21)}, \quad Z^{(43)}\to -Z^{(43)} \, .
\een
On the other hand, for the choice of $\WW_2$ 
without the sign $(-1)^{\delta_{k1} \delta_{\ell 3} \delta_{m4}}$, i.e. all
$Z$-$Z$-$Z$ coupling coming with positive coefficients, this property will be 
lost.\footnote{Since the $c^{(k\ell)}$'s are only determined up to phases in terms
of the background fields, 
one may wonder whether the analysis that led to the conclusion
that the choice of all positive signs in $\WW_2$ is not allowed could
be modified if we use $c^{(k\ell)}$ with different phases. To this end we note
that the information that was needed to arrive at this result was that
under the exchange $x^4\leftrightarrow x^6$ and $x^5\leftrightarrow x^7$, 
$c^{(12)}$ remains unchanged
and $c^{(34)}$ changes sign. These transformation laws of $c^{(12)}$ and
$c^{(34)}$ do not depend on the choice of phases of the $c^{(k\ell)}$'s given 
in \refb{esol2}. Thus our argument
holds.
}
This shows that the form of $\WW_2$ given in \refb{ew2genxx} is the correct
one, and also fixes the relative signs between the terms in $\WW_4$ involving
$\Phi^{(1)}_k$ and $\Phi^{(2)}_k$. 

We now turn to the second exchange symmetry, generated by the
transformation $x^6\leftrightarrow x^8$, 
$x^7\leftrightarrow x^9$ and $N_2\leftrightarrow N_3$.
This exchanges the second and the third D-brane stacks.
\refb{esol2} shows that under this transformation $c^{(14)}\to - c^{(14)}$, 
$c^{(23)}\to c^{(23)}$,
$c^{(12)}\leftrightarrow
c^{(13)}$  and $c^{(24)}\leftrightarrow c^{(34)}$.
We can verify that all the $\WW_i$'s change sign if we accompany these
transformations of $c^{(k\ell)}$ and $N_i$ with the following transformation on the fields:
\ben \label{etr22}
&& (\Phi^{(4)}_1 , \Phi^{(1)}_1) \to (\Phi^{(4)}_1, \Phi^{(1)}_1), \quad 
(\Phi^{(1)}_3 , \Phi^{(2)}_3) \leftrightarrow (\Phi^{(1)}_2,\Phi^{(3)}_2 ), \nonumber \\ && 
(\Phi^{(2)}_1 , \Phi^{(3)}_1) \to (\Phi^{(3)}_1, \Phi^{(2)}_1), \quad
(\Phi^{(4)}_2 , \Phi^{(2)}_2) \leftrightarrow -(\Phi^{(4)}_3, \Phi^{(3)}_3), \nonumber \\
&& Z^{(41)}\to Z^{(41)}, \quad Z^{(2i)} \leftrightarrow -Z^{(3i)},
\quad Z^{(i2)} \leftrightarrow -Z^{(i3)},
\quad \hbox{for $i=1,4$}, \nonumber \\ && 
Z^{(32)}\leftrightarrow -Z^{(23)}, \quad Z^{(14)}\to -Z^{(14)} \, .
\een
It follows from \refb{etr11} and \refb{etr22} that if for $i=1$ and $3$, 
$\Phi^{(i)}_i$ and $\Phi^{(4)}_i$
are interpreted as Wilson lines along the $w^i$ direction on 
the respective branes, then $\Phi^{(2)}_2$ and $\Phi^{(4)}_2$ should
be interpreted as Wilson lines along the $-w^2$ direction on the D2-brane
along the 6-7 directions and the D6-brane, respectively.
This analysis also fixes the relative signs between the terms in $\WW_4$ involving
$\Phi^{(2)}_k$ and $\Phi^{(3)}_k$. 

The theory under consideration also has a symmetry that exchanges 
brane stacks 1 and 4, leaving the stacks 2 and 3 unchanged. This is induced by
making a T-duality transformation 
along 6-7-8-9 directions and then performing an exchange
$6\leftrightarrow 8$, $7\leftrightarrow 9$, and at the same time
exchanging $N_1$ and $N_4$. We shall use the convention that, under a
T-duality transformation along the $x^m$ direction, $g_{mn}\leftrightarrow b_{mn}$
to leading order in $g_{mn}$ and $b_{mn}$ for $n\ne m$ . 
It is easy to see from 
\refb{esol2} 
that under
this exchange $c^{(14)}$ remains unchanged, $c^{(23)}$ changes sign,
$c^{(12)}\leftrightarrow i \, c^{(24)}$ and $c^{(13)}\leftrightarrow i \, c^{(34)}$.
It can be seen that the following accompanying
transformation rules of the fields take $\WW_i$ to $-\WW_i$ for
$1\le i \le 4$:
\ben \label{etr33xx}
&& (\Phi^{(4)}_1 , \Phi^{(1)}_1) \to (\Phi^{(1)}_1, \Phi^{(4)}_1), \quad 
(\Phi^{(1)}_2,\Phi^{(3)}_2 ) 
\leftrightarrow i\, (\Phi^{(4)}_3, \Phi^{(3)}_3), \nonumber \\ && 
(\Phi^{(2)}_1 , \Phi^{(3)}_1) \to (\Phi^{(2)}_1, \Phi^{(3)}_1), \quad
(\Phi^{(4)}_2 , \Phi^{(2)}_2) \leftrightarrow  i\, (\Phi^{(1)}_3,\Phi^{(2)}_3 ), \nonumber \\
&& Z^{(12)}\leftrightarrow  -Z^{(42)},  \quad Z^{(13)}\leftrightarrow -i \, Z^{(43)},
\quad Z^{(21)} \leftrightarrow -i Z^{(24)},
\quad Z^{(31)} \leftrightarrow -Z^{(34)}, \nonumber \\ && 
 Z^{(14)}\leftrightarrow i\, Z^{(41)}, \quad 
 Z^{(32)}\rightarrow Z^{(32)}, \quad Z^{(23)}\to - Z^{(23)} \, .
\een
This fixes the relative signs between the terms in $\WW_4$ involving
$\Phi^{(1)}_k$ and $\Phi^{(4)}_k$.  The factors of $i$ in the transformation
laws relating $\Phi^{(k)}_i$'s to $\Phi^{(k)}_k$'s and $\Phi^{(4)}_k$'s for 
$1\le i,k\le 3$, $i\ne k$ show that if we interpret $\Phi^{(k)}_i$ as the
position of the $k$-th D2-brane along $w^i$, then up to signs,
$i\, \Phi^{(4)}_i$ and
$i\, \Phi^{(i)}_i$ are to be interpreted as  Wilson lines along $w^i$ 
on the $i$-th D2-brane and the D6-brane, respectively. Due to the comments
below \refb{etr22} it follows that there is a further factor
of $-1$ in the definition of $\Phi^{(2)}_2$ and $\Phi^{(4)}_2$ relative to those
for $\Phi^{(i)}_i$ and $\Phi^{(4)}_i$ for $i=1,3$.

All other exchange symmetries
are compositions of the above three transformations and hence
invariance of the potential
under the former follows as
a consequence of their invariance under the latter.
Thus we see that the exchange symmetries 
together with judicious utilization of the field redefinition freedom 
fixes the signs of all the terms in the superpotential except for the overall
sign of $\WW_4$ relative to the other terms.  

The superpotential determined
this way also has other desired symmetries. For example  we see from
\refb{esol2} that under the
transformation $x^4\to x^5$, $x^5\to - x^4$, $c^{(12)}$, $c^{(13)}$, 
$c^{(24)}$ and $c^{(34)}$ get multiplied by $-i$ while the other
$c^{(k\ell)}$'s remain unchanged. It is easy to see that under this 
transformation the superpotential gets multiplied by an overall factor of $-i$,
and hence leaves the potential invariant, if we transform the various fields
as
\ben
&& \Phi^{(k)}_1\to -i\Phi^{(k)}_1 \quad \hbox{for $1\le k\le 4$}, \quad
Z^{(12)}\to -i Z^{(12)},   \quad
Z^{(13)}\to -i Z^{(13)},   
\nonumber \\
&& Z^{(42)}\to -i Z^{(42)},   
\quad
Z^{(43)}\to -i Z^{(43)},
\een  
leaving the other fields invariant.

Finally consider the world-sheet parity transformation under which the NS-NS 2-form field changes
sign. This by itself is not a symmetry of type IIA string theory, but becomes a symmetry if we accompany
this by the parity transformation along the non-compact directions and $(-1)^{F_L}$ -- this is
simply the symmetry group by which we quotient the theory to generate orientifold 6-planes.
Furthermore this transformation leaves the D6- and D2-branes invariant. 
We see from \refb{esol2} that under
this transformation $c^{(k\ell)}\to c^{(k\ell)}$ and $c^{(k4)}\to - c^{(k4)}$ for $1\le k<\ell \le 3$. It is easy to see
that the following transformation on the fields combined with the above 
leaves the superpotential invariant:
\ben \label{etr55xx}
&& \Phi^{(4)}_i \to - (\Phi^{(4)}_i)^T, \quad \Phi^{(i)}_i \to -(\Phi^{(i)}_i)^T, \quad \Phi^{(i)}_j \to (\Phi^{(i)}_j)^T
\quad \hbox{for $1\le i,j\le 3$, $j\ne i$}, \nonumber \\ && 
Z^{(k\ell)} \to (Z^{(\ell k)})^T \quad \hbox{for $1\le k,\ell\le 3$, $\ell\ne k$}, 
\nonumber \\
&&
 Z^{(14)}\rightarrow  (Z^{(41)})^T,  \quad Z^{(24)}\rightarrow - (Z^{(42)})^T, \quad
 Z^{(34)} \to (Z^{(43)})^T, \nonumber \\ && 
Z^{(41)} \rightarrow -(Z^{(14)})^T,
\quad Z^{(42)} \rightarrow (Z^{(24)})^T, \quad
 Z^{(43)}\leftrightarrow -(Z^{(34)})^T \, .
\een
The transposition operation involved in the transformation laws is a reflection of the
fact that under world-sheet parity transformation open strings change their orientation.

Given the superpotential constructed above, the potential $V$ is given by 
\refb{efterm}-\refb{evtotal}. The shift symmetry \refb{eflatgen} generalizes to
\ben \label{eflatgen11}
&& \Phi^{(k)}_m \to \Phi^{(k)}_m+\xi_m I_{N_k},  \quad \hbox{for} 
\quad 1\le k \le 3, \quad k \ne m; \quad  1\le m\le 3, 
\nonumber \\
&& \Phi^{(k)}_k \to \Phi^{(k)}_k + \zeta_k I_{N_k}, \quad \Phi^{(4)}_k \to
\Phi^{(4)}_k+\zeta_k I_{N_4}, \quad \hbox{for} \quad 1\le k\le 3\, , \nonumber \\
&& X^{(k)}_i \to X^{(k)}_i + a_i \, I_{N_k}\, , \quad \hbox{for} \quad 1\le i\le 3\, .
\een

We shall now 
argue that at a generic point in the moduli space,
the vanishing of $V_{gauge}$ given in \refb{egaugegen}
requires all the $X^{(k)}_i$'s to vanish (up to the shift symmetry described in the
last line of \refb{eflatgen11}). To see this first note that the vanishing of $V_{gauge}$
requires each of the terms in \refb{egaugegen} to vanish separately since each is a
positive definite term. The vanishing of the last term tells us that $X^{(k)}_i$ and
$X^{(k)}_j$ commute. Since $X^{(k)}_i$'s are hermitian matrices, this implies that with
the help of $U(N_k)$ gauge transformations we can simultaneously diagonalize each
$X^{(k)}_i$. Therefore we can take
\be \label{edai}
\left(X^{(k)}_i\right)_{mn} = a^{(k)}_{i,m} \, \delta_{mn}\, , \quad \hbox{for} \quad 1\le
m,n\le N_k, \quad 1\le k\le 4\, .
\ee
Physically $a^{(k)}_{i,m}$ has the interpretation of the position along $x^i$ of the 
$m$-th brane in the $k$-th stack.  Now requiring that the first two terms in \refb{egaugegen}
vanish we can easily see that 
\ben 
\left(Z^{(k\ell)}\right)_{mn} \left(a^{(k)}_{i,m} - a^{(\ell)}_{i,n}\right) &=& 0
 \quad \hbox{for $1\le m\le N_k$, $1\le n\le N_\ell$, $1\le i\le 3$}\, ,
\nonumber \\
\left(\Phi^{(k)}\right)_{mn} \left( a^{(k)}_{i,m} - a^{(k)}_{i,n} \right)&=& 0 
 \quad \hbox{for $1\le m,n\le N_k$, $1\le i\le 3$}\, .
\een
This gives
\ben \label{ezph}
\left(Z^{(k\ell)}\right)_{mn} &=& 0 \quad \hbox{if $a^{(k)}_{i,m} \ne a^{(\ell)}_{i,n}$ for any $i$ 
for $1\le m\le N_k$, $1\le n\le N_\ell$}\, ,
\nonumber \\
\left(\Phi^{(k)}\right)_{mn} &=& 0 \quad \hbox{if $a^{(k)}_{i,m} \ne a^{(k)}_{i,n}$ for any $i$
for $1\le m,n\le N_k$}\, .
\een
Furthermore, once \refb{ezph} is satisfied, there is no further constraint on 
the $a^{(k)}_{i,m}$'s.
Physically this means that if the $m$'th brane in the $k$-th stack and the
$n$-th brane in the $\ell$-th stack are separated along the non-compact directions, 
then all fields
associated with the open strings stretched between the two branes must have
zero expectation value. Therefore we can divide the brane stacks into groups, where
within each group all branes are coincident along the non-compact directions, and
the brane stacks in two different groups are separated from each other along at least
one of the non-compact directions. In this case all fields associated with open strings
stretching between two different groups must vanish, and once this condition is
satisfied we have no further constraint on the locations of the groups along the
non-compact directions.
In particular we can separate the different groups by arbitrary distance. 
If such a solution exists, then
the mass of the total brane system will  be given by the sum of the masses of
the individual groups. But at a generic point in the moduli space away from 
subspaces of marginal stability this violates the BPS
bound if the total charge vector is primitive, i.e.\ $\gcd\{N_1,N_2,N_3,N_4\}=1$. 
This shows that it should be impossible to satisfy the F- and D-term equations in this
case. Therefore the only way to solve the equations is to set all the $a^{(k)}_{i,m}$'s
to be equal for any given $i$. Using the shift symmetry given in the last line of
\refb{eflatgen11} we can set these to zero, showing that all the $X^{(k)}_i$'s can
be taken to vanish.

We can also see the absence of solutions with more than one group by directly
analyzing the F-term equations. Let us consider for example the equations we get
by setting the variation of $\WW$ with respect to $\Phi^{(1)}_3$ and $\Phi^{(2)}_3$
to zero. Using \refb{ew1genxx}-\refb{ew4genxx} these equations take the form
\ben \label{esubset}
&& Z^{(12)} Z^{(21)} + N_2 c^{(12)} \, I_{N_1} -[\Phi^{(1)}_1, \Phi^{(1)}_2] = 0\, ,
\nonumber \\
&& Z^{(21)} Z^{(12)} + N_1 c^{(12)} \, I_{N_2} + [\Phi^{(2)}_1, \Phi^{(2)}_2] = 0\, .
\een
Now suppose that we have a solution with multiple groups, with the first group
containing $M_1$ D2-branes along 4-5 directions, $M_2$ D2-branes along 6-7
directions, $M_3$ D2-branes along 8-9 directions and $M_4$ D6-branes along
4-5-6-7-8-9 directions. Let us now take the trace of the first equation over the 
$M_1$ D2-branes in the 4-5 directions and the trace of the second equation over
the $M_2$ D2-branes in the 6-7 directions. Since there are no components of
$Z^{(k\ell)}$ or $\Phi^{(k)}_m$ with one leg in the first group and another leg
in another group, the traces described above restrict the sum to be over the 
indices in the first group only. Denoting the trace with all indices in the first group
by $\hbox{Tr}_1$, we
get
\be \label{esubset1}
\hbox{Tr}_1 (Z^{(12)} Z^{(21)}) + c^{(12)} N_2 M_1 = 0\, ,
\quad \hbox{Tr}_1 (Z^{(21)} Z^{(12)}) + c^{(12)} N_1 M_2 = 0\, .
\ee
Taking the difference between the two equations we see that we must have
$N_1 M_2=N_2 M_1$. Repeating the analysis for other equations we get
$N_i M_j = N_j M_i$ for $1\le i<j\le 4$, 
or equivalently, $M_i/N_i$=constant. The same analysis
may be repeated for other groups, showing that the charge vector of each group
must be proportional to the total charge vector. But this is impossible if the total
charge vector is primitive. This shows that for primitive charge vector there are
no solutions to the F-term equations with more than one group.

For completeness we also give the expected number of solutions to the 
F- and D-term constraints from the counting in the
dual description\cite{0506151}. 
For the $N_i$'s satisfying the restriction
\be 
\gcd\{N_1 N_2, N_1 N_3, N_1 N_4, N_2 N_3, N_2 N_4, N_3 N_4\}=1 \, ,
\ee
this is given by\cite{1405.0412}
\be
-\sum_{s|s_0} s\, \wh c(4 N_1N_2N_3N_4 /s^2)
\ee
where $\wh c(u)$ has been defined in \refb{ek6.5} and
\be \label{es0}
s_0= \prod_{i,j=1\atop i<j}^4\,  \gcd\{N_i, N_j\}\, .
\ee
Note that \refb{es0} is a slight rewriting of the corresponding expression given in 
\cite{1405.0412}.

\sectiono{Explicit solutions for $N=2$} \label{ea1}

In this appendix we shall give the explicit solutions to the F- and D-term
equations for the $N=2$ case. We give the results for the following choice
of parameters:
\ben
&& (c^{(12)}, c^{(13)}, c^{(14)}, c^{(23)}, c^{(24)}, c^{(34)}, c^{(1)}, c^{(2)},
c^{(3)}, c^{(4)}) \nonumber \\
&=& \left(\frac{2}{3},\frac{3}{5}, \frac{5}{7}, \frac{7}{11}, \frac{11}{13}, \frac{13}{17}, \frac{17}{19}, \frac{ 19}{23}, \frac{23}{29}, - \frac{31859}{25346 }  \right) \, .                              
\een
If we scale all the $c^{(k\ell)}$'s and $c^{(k)}$'s by a real positive parameter $\lambda$, then the solutions
given below get scaled by an overall factor of $\sqrt\lambda$. As discussed in footnote \ref{f77}, we can justify the
dropping of higher order terms in the superpotential and the periodicity of the variables $\Phi^{(k)}_i$
if we take the limit $\lambda\to 0$. This is the way we should interpret our solutions, although, in order to avoid 
cluttering, we shall refrain from displaying the factors of $\lambda$ and $\sqrt\lambda$ in the 
expressions for the parameters and solutions, respectively.

In the gauge \refb{egauge1}  we give the 56 solutions below by specifying, for each
solution, the variables in the following order:
\ben 
&& Z^{(12)}, Z^{(21)}, Z^{(13)}, Z^{(31)}, Z^{(23)}, Z^{(32)}, Z^{(14)}_1, 
Z^{(14)}_2,  Z^{(41)}_1, Z^{(41)}_2, Z^{(24)}_1, Z^{(24)}_2, 
 \nnn
Z^{(42)}_1, 
Z^{(42)}_2, Z^{(34)}_1, Z^{(34)}_2, Z^{(43)}_1, Z^{(43)}_2, 
\Phi^{(2)}_3, \Phi^{(3)}_1, \Phi^{(3)}_2, \Phi^{(4)}_{1,11}, \Phi^{(4)}_{1,12},
\Phi^{(4)}_{1,21}, \Phi^{(4)}_{1,22}, 
\nnn\Phi^{(4)}_{2,11}, \Phi^{(4)}_{2,12},
\Phi^{(4)}_{2,21}, \Phi^{(4)}_{2,22},
\Phi^{(4)}_{3,11}, \Phi^{(4)}_{3,12},
\Phi^{(4)}_{3,21}, \Phi^{(4)}_{3,22} \, .
\een
The solutions 
can be organized by two $Z_2$ symmetries. The first $Z_2$ 
corresponds to complex conjugation of all the fields. Since the parameters
$c^{(k\ell)}$ and $c^{(k)}$ have been chosen to be 
real, the F- and D-term constraints are invariant
under complex conjugation and hence given a solution, its complex conjugate 
will also be a solution. 
The second $Z_2$ corresponds to a change in sign of all the
fields under which the superpotential picks up an overall minus sign, and hence 
again the F- and D-term constraints are invariant under this 
transformation.
However since the gauge conditions \refb{egauge2}, \refb{egauge3} and
\refb{egauge4} are not preserved by the second transformation, we have to
accompany this with a compensating gauge transformation to restore the gauge.
This can be done with the help of the element $-1$ of the second U(1) and
the element ${\rm diag}(-1, 1)$ of the U(2) group. It turns out that 12 of the 56
solutions are invariant under the product of the two $Z_2$ transformations -- they
are the first twelve among the solutions listed below. 
These solutions come in pairs related by the first $Z_2$ transformation.
The others transform
non-trivially under both $Z_2$'s and hence come in groups of four, related
by the $Z_2\times Z_2$ transformation. The full set of solutions are:

{\tiny
\ben && 1.1031,-0.60435,0.48800,-1.2295,1.0504,-0.60582,1.3763,-0.63328,-0.73151,0.66599,0,1.4940,0.086853,-1.1327,0.99557,-0.83985,\nnn -1.0579,0.56703, 0.49066,0.52570,1.3450,-0.66999,1.1383,-0.68895,-
   0.77563,0.46310,0.18210,0.14322, -0.33432,-0.75607,-0.62011,\nnn 1.1187,
   -0.58452\, ,
\een 
\ben && 1.1031,-0.60435,-0.48800,1.2295,1.0504,-0.60582,1.3763,0.63328,-0.73151,-0.66599,0,1.4940,-0.086853,-1.1327,0.99557,0.83985,\nnn -1.0579,-0.56703,-0.49066,-0.52570,-1.3450,0.66999,1.1383,-0.6889
   5,0.77563,-0.46310,0.18210,0.14322,0.33432,0.75607,-0.62011,\nnn 1.1187,0.58452
\, , \een    
\ben && 0.61097,-1.0912,1.2046 i,0.49808 i,0.62251,-1.0223,1.4463,0.72947 i,-0.26673,1.4295 i,0,1.4840,0.48529 i,-1.1404,1.0588,-1.3941 i,\nnn -1.3097, -0.10235 i,-2.2283 i,-2.3555 i,-1.1947 i,-0.85898
   i,-0.98702,-0.045414,-0.89763 i,-2.9804 i,0.23714,-0.61928,-1.1212 i,\nnn -0.46046 i,-0.19871, -0.86642,-1.1979 i
\, , \een 
\ben && 0.61097,-1.0912,-1.2046 i,-0.49808 i,0.62251,-1.0223,1.4463,-0.72947 i,-0.26673,-1.4295 i,0,1.4840,-0.48529 i,-1.1404, 1.0588,\nnn 1.3941 i, -1.3097,0.10235 i,2.2283 i,2.3555 i,1.1947 i,0.85898
   i,-0.98702,-0.045414,0.89763 i,2.9804 i,0.23714,-0.61928, 1.1212 i,\nnn  0.46046 i,-0.19871,-0.86642,1.1979 i
\, , \een
\ben && 0.73263,-0.90996,1.1484 i,0.52249 i,0.73592,-0.86472,1.0930,0.72576 i,-0.55912,1.1263 i,0,1.4575,0.18706 i,-1.1611,1.0416, -1.2563 i,\nnn -0.89879,-0.47215 i,-2.2265 i,-2.3550 i,-0.087600
   i,-0.82484 i,-0.97754,-0.40590,-0.95563 i,-2.2487 i,0.045714, -0.15648,\nnn -1.0875 i,-0.67719 i,-0.36232,-1.0160,-1.0046 i
\, , \een
\ben && 0.73263,-0.90996,-1.1484 i,-0.52249 i,0.73592,-0.86472,1.0930,-0.72576 i,-0.55912,-1.1263 i,0,1.4575,-0.18706 i,-1.1611,1.0416,\nnn 1.2563 i, -0.89879,0.47215 i,2.2265 i,2.3550 i,0.087600
   i,0.82484 i,-0.97754,-0.40590,0.95563 i,2.2487 i,0.045714,-0.15648,1.0875 i,\nnn 0.67719 i,-0.36232,-1.0160,1.0046 i 
\, , \een 
\ben && 0.66655,-1.0002,0.75311 i,0.79669 i,0.59925,-1.0619,1.0227,-1.3300 i,0.080251,-1.1358 i,0,1.5175,0.14385 i,-1.1152,-1.1436,\nnn -0.73340 i, 1.3732,0.055941 i,1.2460 i,-1.1153 i,2.4939 i,0.92246
   i,-0.77136,0.061744,0.93839 i,1.9135 i,1.0129,-1.1256,0.58697 i,\nnn 0.0064870 i, 0.30129,-1.1211,-0.28274 i
\, , \een  
\ben && 0.66655,-1.0002,-0.75311 i,-0.79669 i,0.59925,-1.0619,1.0227,1.3300 i,0.080251,1.1358 i,0,1.5175,-0.14385 i,-1.1152,-1.1436,\nnn 0.73340 i, 1.3732,-0.055941 i,-1.2460 i,1.1153 i,-2.4939
    i,-0.92246 i,-0.77136,0.061744,-0.93839 i,-1.9135 i,1.0129,-1.1256,\nnn -0.58697 i,-0.0064870 i,0.30129,-1.1211,0.28274 i 
\, , \een
\ben && 0.68357,-0.97528,-0.78353 i,-0.76577 i,0.69051,-0.92159,2.0781,-0.27085 i,-0.88339,1.5034 i,0,1.7460,-1.1808 i, -0.96926,-0.70695,\nnn -1.8244 i, -1.2693,-1.3301 i,-2.1494 i,-2.2566 i,3.2237
   i,-0.29046 i,1.2418,-0.18345,0.15652 i,1.0460 i,0.69047,-1.4404,-0.57024 i,\nnn  0.67798 i,0.48642,0.60953,-0.80716 i
\, , \een 
\ben && 0.68357,-0.97528,0.78353 i,0.76577 i,0.69051,-0.92159,2.0781,0.27085 i,-0.88339,-1.5034 i,0,1.7460,1.1808 i,-0.96926, -0.70695,\nnn 1.8244 i, -1.2693,1.3301 i,2.1494 i,2.2566 i,-3.2237 i,0.29046
   i,1.2418,-0.18345,-0.15652 i,-1.0460 i,0.69047,-1.4404, 0.57024 i,\nnn -0.67798 i, 0.48642,0.60953,0.80716 i
\, , \een
\ben && 0.59598,-1.1186,-0.73969 i,-0.81115 i,0.65823,-0.96678,1.6985,0.30140 i,-0.92427,-0.46869 i,0,1.3919,0.16684 i,-1.2158,0.29842,\nnn -1.4089 i, 0.63575,-1.2202 i,1.0605 i,-1.3233 i,-2.4873
    i,0.19243 i,1.1190,-0.35198,0.095824 i,-1.6071 i,1.1791,-1.2238,-0.90847 i,\nnn -0.98383 i,-0.042304,0.76948,-0.77265 i 
\, , \een 
\ben && 0.59598,-1.1186,0.73969 i,0.81115 i,0.65823,-0.96678,1.6985,-0.30140 i,-0.92427,0.46869 i,0,1.3919,-0.16684 i,-1.2158,0.29842,\nnn 1.4089 i,0.63575,1.2202 i,-1.0605 i,1.3233 i,2.4873 i,-0.19243
   i,1.1190,-0.35198,-0.095824 i,1.6071 i,1.1791,-1.2238,0.90847 i,\nnn 0.98383 i,  -0.042304,0.76948,0.77265 i
\, , \een
\ben && 0.45313,-1.4712,0.56822-0.90538 i,-0.29838-0.47544 i,0.44418,-1.4327,1.6133,-0.28628+0.85922 i,-1.06265-0.20046 i,\nnn 0.23903-0.41227 i,0,1.4603,0.25762-0.03156 i,-1.1589, -1.18478-0.54983
   i,-0.58503+0.44603 i,1.29014-0.03809 i,\nnn 0.52573-0.73456 i, -0.14716+0.55274 i, -0.15369+0.58393 i,-0.8983-2.4883 i,0.64477+0.07407 i,0.51341-0.56880 i,\nnn -0.96725+0.17004 i,0.22399+0.20236
   i, -0.12969-0.36193 i,1.9718-0.0456 i,-1.9858-0.1672 i,0.11047-0.72999 i,\nnn 0.50307+0.02573 i,0.94973-0.33977 i, -0.48769+0.41436 i,0.30881+0.23653 i
\, , \een
\ben && 0.45313,-1.4712,0.56822+0.90538 i,-0.29838+0.47544 i,0.44418,-1.4327,1.6133,-0.28628-0.85922 i,-1.06265+0.20046 i,\nnn 0.23903+0.41227 i,0,1.4603,0.25762+0.03156 i,-1.1589,-1.18478+0.54983
   i,-0.58503-0.44603 i,1.29014+0.03809 i,\nnn 0.52573+0.73456 i,-0.14716-0.55274 i,-0.15369-0.58393 i,-0.8983+2.4883 i,0.64477-0.07407 i,0.51341+0.56880 i,\nnn -0.96725-0.17004 i,0.22399-0.20236
   i,-0.12969+0.36193 i,1.9718+0.0456 i,-1.9858+0.1672 i,0.11047+0.72999 i,\nnn 0.50307-0.02573 i,0.94973+0.33977 i,-0.48769-0.41436 i,0.30881-0.23653 i
\, , \een
\ben && 0.45313,-1.4712,-0.56822+0.90538 i,0.29838+0.47544 i,0.44418,-1.4327,1.6133,0.28628-0.85922 i,-1.06265-0.20046 i,\nnn -0.23903+0.41227 i,0,1.4603,-0.25762+0.03156 i,-1.1589,-1.18478-0.54983
   i,0.58503-0.44603 i,1.29014-0.03809 i,\nnn -0.52573+0.73456 i,0.14716-0.55274 i,0.15369-0.58393 i,0.8983+2.4883 i,-0.64477-0.07407 i,0.51341-0.56880 i,\nnn  -0.96725+0.17004 i,-0.22399-0.20236
   i,0.12969+0.36193 i,1.9718-0.0456 i,-1.9858-0.1672 i,-0.11047+0.72999 i,\nnn -0.50307-0.02573 i,0.94973-0.33977 i, -0.48769+0.41436 i,-0.30881-0.23653 i
\, , \een
\ben && 0.45313,-1.4712,-0.56822-0.90538 i,0.29838-0.47544 i,0.44418,-1.4327,1.6133,0.28628+0.85922 i,-1.06265+0.20046 i,\nnn -0.23903-0.41227 i,0,1.4603,-0.25762-0.03156 i,-1.1589,-1.18478+0.54983
   i,0.58503+0.44603 i,1.29014+0.03809 i,\nnn -0.52573-0.73456 i,0.14716+0.55274 i,0.15369+0.58393 i,0.8983-2.4883 i,-0.64477+0.07407 i,0.51341+0.56880 i,\nnn -0.96725-0.17004 i,-0.22399+0.20236
   i,0.12969-0.36193 i,1.9718+0.0456 i,-1.9858+0.1672 i,-0.11047-0.72999 i,\nnn -0.50307+0.02573 i,0.94973+0.33977 i,-0.48769-0.41436 i,-0.30881+0.23653 i
\, , \een
\ben && 0.95002,-0.70174,-0.32710+0.46207 i,0.61236+0.86502 i,0.91139,-0.69823,1.4042,0.69607+0.01764 i,-1.07310+0.03164 i,\nnn 0.110778-0.066634 i,0,1.5414,-0.32109-0.41749 i,-1.0979,0.55005+1.21024
   i,0.16063-0.43612 i,-0.57733+0.76664 i,\nnn -0.77052-0.36748 i,-1.0386-0.9771 i,-1.0513-0.9773 i,-1.6481+0.1394 i,0.75969-0.23579 i,0.16464-0.02762 i,\nnn -0.45318+0.69073 i,1.5501-0.2952
   i,0.21242-0.81845 i,0.18804+0.61876 i,-0.31405+0.71559 i,-0.22192-0.26590 i,\nnn 1.13165-0.17765 i,0.54413+0.83642 i,0.132118+0.002732 i,1.05645-0.27973 i
\, , \een
\ben && 0.95002,-0.70174,-0.32710-0.46207 i,0.61236-0.86502 i,0.91139,-0.69823,1.4042,0.69607-0.01764 i,-1.07310-0.03164 i,\nnn 0.110778+0.066634 i,0,1.5414,-0.32109+0.41749 i,-1.0979,0.55005-1.21024
    i,0.16063+0.43612 i,-0.57733-0.76664 i,\nnn -0.77052+0.36748 i,-1.0386+0.9771 i,-1.0513+0.9773 i,-1.6481-0.1394 i,0.75969+0.23579 i,0.16464+0.02762 i,\nnn -0.45318-0.69073 i,1.5501+0.2952
    i,0.21242+0.81845 i,0.18804-0.61876 i,-0.31405-0.71559 i,-0.22192+0.26590 i,\nnn 1.13165+0.17765 i,0.54413-0.83642 i,0.132118-0.002732 i,1.05645+0.27973 i
\, , \een
\ben && 0.95002,-0.70174,0.32710-0.46207 i,-0.61236-0.86502 i,0.91139,-0.69823,1.4042,-0.69607-0.01764 i,-1.07310+0.03164 i,\nnn -0.110778+0.066634 i,0,1.5414,0.32109+0.41749 i,-1.0979,0.55005+1.21024
   i,-0.16063+0.43612 i,-0.57733+0.76664 i,\nnn 0.77052+0.36748 i,1.0386+0.9771 i,1.0513+0.9773 i,1.6481-0.1394 i,-0.75969+0.23579 i,0.16464-0.02762 i, \nnn -0.45318+0.69073 i,-1.5501+0.2952
   i,-0.21242+0.81845 i,0.18804+0.61876 i,-0.31405+0.71559 i,0.22192+0.26590 i,\nnn -1.13165+0.17765 i,0.54413+0.83642 i,0.132118+0.002732 i,-1.05645+0.27973 i
\, , \een
\ben && 0.95002,-0.70174,0.32710+0.46207 i,-0.61236+0.86502 i,0.91139,-0.69823,1.4042,-0.69607+0.01764 i,-1.07310-0.03164 i,\nnn -0.110778-0.066634 i,0,1.5414,0.32109-0.41749 i,-1.0979,0.55005-1.21024
   i,-0.16063-0.43612 i,-0.57733-0.76664 i,\nnn 0.77052-0.36748 i,1.0386-0.9771 i,1.0513-0.9773 i,1.6481+0.1394 i,-0.75969-0.23579 i,0.16464+0.02762 i,\nnn -0.45318-0.69073 i,-1.5501-0.2952
   i,-0.21242-0.81845 i,0.18804-0.61876 i,-0.31405-0.71559 i,0.22192-0.26590 i,\nnn -1.13165-0.17765 i,0.54413-0.83642 i,0.132118-0.002732 i,-1.05645-0.27973 i
\, , \een
\ben && 0.91406,-0.72935,-0.32785+0.21228 i,1.2895+0.8349 i,1.4439,-0.44072,1.9777,0.59874-0.39052 i,-0.71679+0.35810 i, \nnn 0.52839-0.83821 i,0,1.5872,-1.4616-0.0841 i,-1.0662,0.53349+0.85291
   i,0.25457-0.94589 i,-0.54140+0.29559 i,\nnn -0.56202-0.89376 i,-3.7027+0.1711 i,-0.45217-1.08965 i,-1.31223-0.02937 i,0.28581-0.03529 i,0.28570+0.19367 i,\nnn -0.42571+0.11678 i,1.4713-0.2882
   i,0.25631-0.43236 i,0.00524+0.38763 i,-0.42348+0.77593 i,-0.04355-0.68106 i,\nnn 3.3594-0.7888 i,1.36463+0.09310 i,0.80459-0.12635 i,1.13364-0.56938 i
\, , \een
\ben && 0.91406,-0.72935,-0.32785-0.21228 i,1.2895-0.8349 i,1.4439,-0.44072,1.9777,0.59874+0.39052 i,-0.71679-0.35810 i,\nnn 0.52839+0.83821 i,0,1.5872,-1.4616+0.0841 i,-1.0662,0.53349-0.85291
   i,0.25457+0.94589 i,-0.54140-0.29559 i,\nnn -0.56202+0.89376 i,-3.7027-0.1711 i,-0.45217+1.08965 i,-1.31223+0.02937 i,0.28581+0.03529 i,0.28570-0.19367 i,\nnn -0.42571-0.11678 i,1.4713+0.2882
   i,0.25631+0.43236 i,0.00524-0.38763 i,-0.42348-0.77593 i,-0.04355+0.68106 i,\nnn 3.3594+0.7888 i,1.36463-0.09310 i,0.80459+0.12635 i,1.13364+0.56938 i
\, , \een  
\ben && 0.91406,-0.72935,0.32785-0.21228 i,-1.2895-0.8349 i,1.4439,-0.44072,1.9777,-0.59874+0.39052 i,-0.71679+0.35810 i,\nnn -0.52839+0.83821 i,0,1.5872,1.4616+0.0841 i,-1.0662,0.53349+0.85291
   i,-0.25457+0.94589 i,-0.54140+0.29559 i,\nnn 0.56202+0.89376 i,3.7027-0.1711 i,0.45217+1.08965 i,1.31223+0.02937 i,-0.28581+0.03529 i,0.28570+0.19367 i,\nnn -0.42571+0.11678 i,-1.4713+0.2882
   i, -0.25631+0.43236 i,0.00524+0.38763 i,-0.42348+0.77593 i,0.04355+0.68106 i,\nnn -3.3594+0.7888 i,1.36463+0.09310 i,0.80459-0.12635 i,-1.13364+0.56938 i
\, , \een 
\ben && 0.91406,-0.72935,0.32785+0.21228 i,-1.2895+0.8349 i,1.4439,-0.44072,1.9777,-0.59874-0.39052 i,-0.71679-0.35810 i,\nnn -0.52839-0.83821 i,0,1.5872,1.4616-0.0841 i,-1.0662,0.53349-0.85291
   i,-0.25457-0.94589 i,-0.54140-0.29559 i,\nnn 0.56202-0.89376 i,3.7027+0.1711 i,0.45217-1.08965 i,1.31223-0.02937 i,-0.28581-0.03529 i,0.28570-0.19367 i,\nnn -0.42571-0.11678 i,-1.4713-0.2882
   i,-0.25631-0.43236 i,0.00524-0.38763 i,-0.42348-0.77593 i,0.04355-0.68106 i,\nnn -3.3594-0.7888 i,1.36463-0.09310 i,0.80459+0.12635 i,-1.13364-0.56938 i
\, , \een 
\ben && 1.6191,-0.41174,-0.31539+0.21907 i,1.2832+0.8914 i,0.89136,-0.71393,1.1087,0.83973+0.29799 i,-1.00374+0.24552 i,\nnn -0.43614-0.16939 i,0,2.2206,-0.24786-1.13810 i,-0.76211,0.9500+1.2734
   i,0.22293+0.17133 i,-0.27088+0.66960 i,\nnn -1.8137+0.0879 i,-0.42589-1.01614 i,-4.3726+0.1178 i,-1.27855-0.01388 i,0.94073-0.59690 i,-0.10924-0.54527 i,\nnn -0.17013+1.07890 i,4.2738-0.7905
   i,0.27296-1.10597 i,0.13837+0.05770 i,0.17575+0.51118 i,-0.066218+0.013520 i,\nnn 1.05447-0.31776 i,0.18575+0.84116 i,-0.06819-0.18417 i,0.54875+0.00570 i
\, , \een
\ben && 1.6191,-0.41174,-0.31539-0.21907 i,1.2832-0.8914 i,0.89136,-0.71393,1.1087,0.83973-0.29799 i,-1.00374-0.24552 i,\nnn -0.43614+0.16939 i,0,2.2206,-0.24786+1.13810 i,-0.76211,0.9500-1.2734
   i,0.22293-0.17133 i,-0.27088-0.66960 i,\nnn -1.8137-0.0879 i,-0.42589+1.01614 i,-4.3726-0.1178 i,-1.27855+0.01388 i,0.94073+0.59690 i,-0.10924+0.54527 i,\nnn -0.17013-1.07890 i,4.2738+0.7905
   i,0.27296+1.10597 i,0.13837-0.05770 i,0.17575-0.51118 i,-0.066218-0.013520 i,\nnn 1.05447+0.31776 i,0.18575-0.84116 i,-0.06819+0.18417 i,0.54875-0.00570 i
\, , \een
\ben && 1.6191,-0.41174,0.31539-0.21907 i,-1.2832-0.8914 i,0.89136,-0.71393,1.1087,-0.83973-0.29799 i,-1.00374+0.24552 i,\nnn 0.43614+0.16939 i,0,2.2206,0.24786+1.13810 i,-0.76211,0.9500+1.2734
   i,-0.22293-0.17133 i,-0.27088+0.66960 i,\nnn 1.8137-0.0879 i,0.42589+1.01614 i,4.3726-0.1178 i,1.27855+0.01388 i,-0.94073+0.59690 i,-0.10924-0.54527 i,\nnn -0.17013+1.07890 i,-4.2738+0.7905
   i,-0.27296+1.10597 i,0.13837+0.05770 i,0.17575+0.51118 i,0.066218-0.013520 i,\nnn -1.05447+0.31776 i,0.18575+0.84116 i,-0.06819-0.18417 i,-0.54875-0.00570 i
\, , \een 
\ben && 1.6191,-0.41174,0.31539+0.21907 i,-1.2832+0.8914 i,0.89136,-0.71393,1.1087,-0.83973+0.29799 i,-1.00374-0.24552 i,\nnn 0.43614-0.16939 i,0,2.2206,0.24786-1.13810 i,-0.76211,0.9500-1.2734
   i,-0.22293+0.17133 i,-0.27088-0.66960 i,\nnn 1.8137+0.0879 i,0.42589-1.01614 i,4.3726+0.1178 i,1.27855-0.01388 i,-0.94073-0.59690 i,-0.10924+0.54527 i,\nnn -0.17013-1.07890 i,-4.2738-0.7905
   i,-0.27296-1.10597 i,0.13837-0.05770 i,0.17575-0.51118 i,0.066218+0.013520 i,\nnn -1.05447-0.31776 i,0.18575-0.84116 i,-0.06819+0.18417 i,-0.54875+0.00570 i
\, , \een
\ben && 1.9767,-0.33727,0.05613-0.20931 i,-0.7171-2.6743 i,1.8266,-0.34838,2.5896,-0.97646+0.06703 i,-0.50833+0.57791 i,\nnn 0.0097+1.5333 i,0,3.0381,2.6956+0.4993 i,-0.55702,0.2200+1.8898
   i,-0.39511+0.29393 i,-0.34960+0.14642 i,\nnn 2.6774+0.4012 i,3.7967+0.6721 i,4.4427+0.7059 i,0.66723-0.48517 i,-0.31466+0.03542 i,0.56011+0.26856 i,\nnn -0.42542+0.12618 i,-4.6583+0.4940
   i,-0.25449+0.46452 i,-0.062714+0.060676 i,-0.15517+1.13620 i,-0.12916+0.16928 i,\nnn -3.7951+0.4902 i,1.3886+0.3523 i,0.32390-0.11002 i,-0.46294+0.00590 i
\, , \een
\ben && 1.9767,-0.33727,0.05613+0.20931 i,-0.7171+2.6743 i,1.8266,-0.34838,2.5896,-0.97646-0.06703 i,-0.50833-0.57791 i,\nnn 0.0097-1.5333 i,0,3.0381,2.6956-0.4993 i,-0.55702,0.2200-1.8898
   i,-0.39511-0.29393 i,-0.34960-0.14642 i,\nnn 2.6774-0.4012 i,3.7967-0.6721 i,4.4427-0.7059 i,0.66723+0.48517 i,-0.31466-0.03542 i,0.56011-0.26856 i,\nnn -0.42542-0.12618 i,-4.6583-0.4940
   i,-0.25449-0.46452 i,-0.062714-0.060676 i,-0.15517-1.13620 i,-0.12916-0.16928 i,\nnn -3.7951-0.4902 i,1.3886-0.3523 i,0.32390+0.11002 i,-0.46294-0.00590 i
\, , \een 
\ben && 1.9767,-0.33727,-0.05613+0.20931 i,0.7171+2.6743 i,1.8266,-0.34838,2.5896,0.97646-0.06703 i,-0.50833+0.57791 i,\nnn -0.0097-1.5333 i,0,3.0381,-2.6956-0.4993 i,-0.55702,0.2200+1.8898
   i,0.39511-0.29393 i,-0.34960+0.14642 i,\nnn -2.6774-0.4012 i,-3.7967-0.6721 i,-4.4427-0.7059 i,-0.66723+0.48517 i,0.31466-0.03542 i,0.56011+0.26856 i,\nnn -0.42542+0.12618 i,4.6583-0.4940
   i,0.25449-0.46452 i,-0.062714+0.060676 i,-0.15517+1.13620 i,0.12916-0.16928 i,\nnn 3.7951-0.4902 i,1.3886+0.3523 i,0.32390-0.11002 i,0.46294-0.00590 i
\, , \een 
\ben && 1.9767,-0.33727,-0.05613-0.20931 i,0.7171-2.6743 i,1.8266,-0.34838,2.5896,0.97646+0.06703 i,-0.50833-0.57791 i,\nnn -0.0097+1.5333 i,0,3.0381,-2.6956+0.4993 i,-0.55702,0.2200-1.8898
   i,0.39511+0.29393 i,-0.34960-0.14642 i,\nnn -2.6774+0.4012 i,-3.7967+0.6721 i,-4.4427+0.7059 i,-0.66723-0.48517 i,0.31466+0.03542 i,0.56011-0.26856 i,\nnn -0.42542-0.12618 i,4.6583+0.4940
   i,0.25449+0.46452 i,-0.062714-0.060676 i,-0.15517-1.13620 i,0.12916+0.16928 i,\nnn 3.7951+0.4902 i,1.3886-0.3523 i,0.32390+0.11002 i,0.46294+0.00590 i
\, , \een
\ben && 1.9110,-0.34887,0.00327-0.22368 i,-0.0393-2.6819 i,1.7784,-0.35783,2.6804,-0.53916+0.27523 i,-0.27538+0.28432 i,\nnn 0.4435+1.6399 i,0,3.0545,-2.7364-0.4622 i,-0.55405,0.2485-1.8820
   i,-0.15028-0.62950 i,-0.84974-0.35066 i,\nnn 2.5088-0.4469 i,-3.6825-0.6863 i,4.3083-0.7236 i,-0.03074+0.48708 i,0.24037-0.02620 i,1.36785+0.30507 i,\nnn 0.42890-0.02582 i,-4.0393-0.4868
   i,0.03734-0.23398 i,0.021371-0.082685 i,-0.16148-1.09578 i,-0.02592-0.39795 i,\nnn 3.6366-0.4677 i,-0.64660-0.11788 i,0.38877-0.11146 i,-0.38953-0.01540 i
\, , \een
\ben && 1.9110,-0.34887,0.00327+0.22368 i,-0.0393+2.6819 i,1.7784,-0.35783,2.6804,-0.53916-0.27523 i,-0.27538-0.28432 i,\nnn 0.4435-1.6399 i,0,3.0545,-2.7364+0.4622 i,-0.55405,0.2485+1.8820
   i,-0.15028+0.62950 i,-0.84974+0.35066 i,\nnn 2.5088+0.4469 i,-3.6825+0.6863 i,4.3083+0.7236 i,-0.03074-0.48708 i,0.24037+0.02620 i,1.36785-0.30507 i,\nnn 0.42890+0.02582 i,-4.0393+0.4868
   i,0.03734+0.23398 i,0.021371+0.082685 i,-0.16148+1.09578 i,-0.02592+0.39795 i,\nnn 3.6366+0.4677 i,-0.64660+0.11788 i,0.38877+0.11146 i,-0.38953+0.01540 i
\, , \een 
\ben && 1.9110,-0.34887,-0.00327+0.22368 i,0.0393+2.6819 i,1.7784,-0.35783,2.6804,0.53916-0.27523 i,-0.27538+0.28432 i,\nnn -0.4435-1.6399 i,0,3.0545,2.7364+0.4622 i,-0.55405,0.2485-1.8820
   i,0.15028+0.62950 i,-0.84974-0.35066 i,\nnn -2.5088+0.4469 i,3.6825+0.6863 i,-4.3083+0.7236 i,0.03074-0.48708 i,-0.24037+0.02620 i,1.36785+0.30507 i,\nnn 0.42890-0.02582 i,4.0393+0.4868
   i,-0.03734+0.23398 i,0.021371-0.082685 i,-0.16148-1.09578 i,0.02592+0.39795 i,\nnn -3.6366+0.4677 i,-0.64660-0.11788 i,0.38877-0.11146 i,0.38953+0.01540 i
\, , \een
\ben && 1.9110,-0.34887,-0.00327-0.22368 i,0.0393-2.6819 i,1.7784,-0.35783,2.6804,0.53916+0.27523 i,-0.27538-0.28432 i,\nnn -0.4435+1.6399 i,0,3.0545,2.7364-0.4622 i,-0.55405,0.2485+1.8820
   i,0.15028-0.62950 i,-0.84974+0.35066 i,\nnn -2.5088-0.4469 i,3.6825-0.6863 i,-4.3083-0.7236 i,0.03074+0.48708 i,-0.24037-0.02620 i,1.36785-0.30507 i,\nnn 0.42890+0.02582 i,4.0393-0.4868
   i,-0.03734-0.23398 i,0.021371+0.082685 i,-0.16148+1.09578 i,0.02592-0.39795 i,\nnn -3.6366-0.4677 i,-0.64660+0.11788 i,0.38877+0.11146 i,0.38953-0.01540 i
\, , \een    
\ben && 1.4754,-0.45185,0.26993-0.31443 i,-0.9431-1.0986 i,0.90647,-0.70202,1.1900,-0.64403+0.44663 i,-0.70679+0.19227 i,\nnn 0.44962+0.66707 i,0,2.1641,-0.58808-1.12062 i,-0.78201,-0.5172-1.5098
   i,-0.13148-0.42354 i,-0.15008-0.87223 i,\nnn 1.6534-0.1717 i,-0.26139-0.99093 i,3.8126-0.2532 i,-0.61422-0.01569 i,0.40219-0.56716 i,0.78585+1.09294 i,\nnn 0.52651-1.06313 i,-3.4366-0.4772
   i,-0.20918-0.81249 i,0.19179-0.23502 i,-0.46509-0.63244 i,0.07940-0.27067 i,\nnn 0.96433-0.28726 i,0.02610-0.29483 i,0.12662-0.40894 i,-0.39814-0.03508 i
\, , \een
\ben && 1.4754,-0.45185,0.26993+0.31443 i,-0.9431+1.0986 i,0.90647,-0.70202,1.1900,-0.64403-0.44663 i,-0.70679-0.19227 i,\nnn 0.44962-0.66707 i,0,2.1641,-0.58808+1.12062 i,-0.78201,-0.5172+1.5098
   i,-0.13148+0.42354 i,-0.15008+0.87223 i,\nnn 1.6534+0.1717 i,-0.26139+0.99093 i,3.8126+0.2532 i,-0.61422+0.01569 i,0.40219+0.56716 i,0.78585-1.09294 i,\nnn 0.52651+1.06313 i,-3.4366+0.4772
   i,-0.20918+0.81249 i,0.19179+0.23502 i,-0.46509+0.63244 i,0.07940+0.27067 i,\nnn 0.96433+0.28726 i,0.02610+0.29483 i,0.12662+0.40894 i,-0.39814+0.03508 i
\, , \een 
\ben && 1.4754,-0.45185,-0.26993+0.31443 i,0.9431+1.0986 i,0.90647,-0.70202,1.1900,0.64403-0.44663 i,-0.70679+0.19227 i,\nnn -0.44962-0.66707 i,0,2.1641,0.58808+1.12062 i,-0.78201,-0.5172-1.5098
   i,0.13148+0.42354 i,-0.15008-0.87223 i,\nnn -1.6534+0.1717 i,0.26139+0.99093 i,-3.8126+0.2532 i,0.61422+0.01569 i,-0.40219+0.56716 i,0.78585+1.09294 i,\nnn 0.52651-1.06313 i,3.4366+0.4772
   i,0.20918+0.81249 i,0.19179-0.23502 i,-0.46509-0.63244 i,-0.07940+0.27067 i,\nnn -0.96433+0.28726 i,0.02610-0.29483 i,0.12662-0.40894 i,0.39814+0.03508 i
\, , \een 
\ben && 1.4754,-0.45185,-0.26993-0.31443 i,0.9431-1.0986 i,0.90647,-0.70202,1.1900,0.64403+0.44663 i,-0.70679-0.19227 i,\nnn -0.44962+0.66707 i,0,2.1641,0.58808-1.12062 i,-0.78201,-0.5172+1.5098
   i,0.13148-0.42354 i,-0.15008+0.87223 i,\nnn-1.6534-0.1717 i,0.26139-0.99093 i,-3.8126-0.2532 i,0.61422-0.01569 i,-0.40219-0.56716 i,0.78585-1.09294 i,\nnn 0.52651+1.06313 i,3.4366-0.4772
   i,0.20918-0.81249 i,0.19179+0.23502 i,-0.46509+0.63244 i,-0.07940-0.27067 i,\nnn -0.96433-0.28726 i,0.02610+0.29483 i,0.12662+0.40894 i,0.39814-0.03508 i
\, , \een
\ben && 0.91505,-0.72856,0.27183+0.31858 i,-0.9300+1.0899 i,1.3057,-0.48737,2.0032,-0.14161-0.38149 i,-0.57789-0.19254 i,\nnn -0.65685-0.95410 i,0,1.6661,1.4419-0.0115 i,-1.0157,0.58147-0.19814
   i,-0.13423+1.20011 i,-0.70261+0.14143 i,\nnn -0.07650+0.96588 i,3.1688+0.2287 i,-0.24929+1.05474 i,-0.57752+0.03039 i,-0.114872-0.004834 i,0.95170+0.00885 i,\nnn 0.27274+0.12463 i,0.66228+0.34284
   i,-0.091636-0.007210 i,0.05332+0.23213 i,-0.42025-0.15528 i,-0.04327+1.10732 i,\nnn -2.8851+0.3857 i,0.92607-0.07941 i,1.16570+0.08767 i,0.42643+0.60833 i
\, , \een 
\ben && 0.91505,-0.72856,0.27183-0.31858 i,-0.9300-1.0899 i,1.3057,-0.48737,2.0032,-0.14161+0.38149 i,-0.57789+0.19254 i,\nnn -0.65685+0.95410 i,0,1.6661,1.4419+0.0115 i,-1.0157,0.58147+0.19814
   i,-0.13423-1.20011 i,-0.70261-0.14143 i,\nnn -0.07650-0.96588 i,3.1688-0.2287 i,-0.24929-1.05474 i,-0.57752-0.03039 i,-0.114872+0.004834 i,0.95170-0.00885 i,\nnn 0.27274-0.12463 i,0.66228-0.34284
   i,-0.091636+0.007210 i,0.05332-0.23213 i,-0.42025+0.15528 i,-0.04327-1.10732 i,\nnn -2.8851-0.3857 i,0.92607+0.07941 i,1.16570-0.08767 i,0.42643-0.60833 i
\, , \een 
\ben && 0.91505,-0.72856,-0.27183-0.31858 i,0.9300-1.0899 i,1.3057,-0.48737,2.0032,0.14161+0.38149 i,-0.57789-0.19254 i,\nnn 0.65685+0.95410 i,0,1.6661,-1.4419+0.0115 i,-1.0157,0.58147-0.19814
   i,0.13423-1.20011 i,-0.70261+0.14143 i,\nnn 0.07650-0.96588 i,-3.1688-0.2287 i,0.24929-1.05474 i,0.57752-0.03039 i,0.114872+0.004834 i,0.95170+0.00885 i,\nnn 0.27274+0.12463 i,-0.66228-0.34284
   i,0.091636+0.007210 i,0.05332+0.23213 i,-0.42025-0.15528 i,0.04327-1.10732 i,\nnn 2.8851-0.3857 i,0.92607-0.07941 i,1.16570+0.08767 i,-0.42643-0.60833 i
\, , \een
\ben && 0.91505,-0.72856,-0.27183+0.31858 i,0.9300+1.0899 i,1.3057,-0.48737,2.0032,0.14161-0.38149 i,-0.57789+0.19254 i,\nnn 0.65685-0.95410 i,0,1.6661,-1.4419-0.0115 i,-1.0157,0.58147+0.19814
    i,0.13423+1.20011 i,-0.70261-0.14143 i,\nnn 0.07650+0.96588 i,-3.1688+0.2287 i,0.24929+1.05474 i,0.57752+0.03039 i,0.114872-0.004834 i,0.95170-0.00885 i,\nnn 0.27274-0.12463 i,-0.66228+0.34284
    i,0.091636-0.007210 i,0.05332-0.23213 i,-0.42025+0.15528 i,0.04327+1.10732 i,\nnn 2.8851+0.3857 i,0.92607+0.07941 i,1.16570-0.08767 i,-0.42643+0.60833 i
\, , \een 
\ben && 0.83084,-0.80240,0.54517+0.28351 i,-0.86630+0.45050 i,0.82640,-0.77005,1.2825,-0.63500-0.65516 i,-0.62821-0.25417 i,\nnn 0.21863-0.73889 i,0,1.4650,0.07479+0.15346 i,-1.1552,0.19925+0.77645
   i,-0.39963+1.01146 i,-0.55839+0.66064 i,\nnn 0.56405+0.67208 i,0.49305+0.88504 i,0.47057+0.91614 i,-0.114229-0.033999 i,-0.03362+0.17079 i,1.07828-0.23571 i,\nnn 0.43204+0.65475 i,-0.15165+0.37036
   i,-0.00587+0.19230 i,0.05368+0.63486 i,-0.59003+0.43800 i,0.12237+0.92941 i,\nnn -0.073227+0.088523 i,0.72886+0.09120 i,0.87484+0.67036 i,-0.13807+0.40776 i
\, , \een
\ben && 0.83084,-0.80240,0.54517-0.28351 i,-0.86630-0.45050 i,0.82640,-0.77005,1.2825,-0.63500+0.65516 i,-0.62821+0.25417 i,\nnn 0.21863+0.73889 i,0,1.4650,0.07479-0.15346 i,-1.1552,0.19925-0.77645
   i,-0.39963-1.01146 i,-0.55839-0.66064 i,\nnn 0.56405-0.67208 i,0.49305-0.88504 i,0.47057-0.91614 i,-0.114229+0.033999 i,-0.03362-0.17079 i,1.07828+0.23571 i,\nnn 0.43204-0.65475 i,-0.15165-0.37036
   i,-0.00587-0.19230 i,0.05368-0.63486 i,-0.59003-0.43800 i,0.12237-0.92941 i,\nnn -0.073227-0.088523 i,0.72886-0.09120 i,0.87484-0.67036 i,-0.13807-0.40776 i
\, , \een    
\ben && 0.83084,-0.80240,-0.54517-0.28351 i,0.86630-0.45050 i,0.82640,-0.77005,1.2825,0.63500+0.65516 i,-0.62821-0.25417 i,\nnn -0.21863+0.73889 i,0,1.4650,-0.07479-0.15346 i,-1.1552,0.19925+0.77645
   i,0.39963-1.01146 i,-0.55839+0.66064 i,\nnn -0.56405-0.67208 i,-0.49305-0.88504 i,-0.47057-0.91614 i,0.114229+0.033999 i,0.03362-0.17079 i,1.07828-0.23571 i,\nnn 0.43204+0.65475 i,0.15165-0.37036
   i,0.00587-0.19230 i,0.05368+0.63486 i,-0.59003+0.43800 i,-0.12237-0.92941 i,\nnn 0.073227-0.088523 i,0.72886+0.09120 i,0.87484+0.67036 i,0.13807-0.40776 i
\, , \een
\ben && 0.83084,-0.80240,-0.54517+0.28351 i,0.86630+0.45050 i,0.82640,-0.77005,1.2825,0.63500-0.65516 i,-0.62821+0.25417 i,\nnn -0.21863-0.73889 i,0,1.4650,-0.07479+0.15346 i,-1.1552,0.19925-0.77645
   i,0.39963+1.01146 i,-0.55839-0.66064 i,\nnn -0.56405+0.67208 i,-0.49305+0.88504 i,-0.47057+0.91614 i,0.114229-0.033999 i,0.03362+0.17079 i,1.07828+0.23571 i,\nnn 0.43204-0.65475 i,0.15165+0.37036
   i,0.00587+0.19230 i,0.05368-0.63486 i,-0.59003-0.43800 i,-0.12237+0.92941 i,\nnn 0.073227+0.088523 i,0.72886-0.09120 i,0.87484-0.67036 i,0.13807+0.40776 i
\, , \een   
\ben && 0.39164,-1.7022,-0.35103-0.95729 i,0.20259-0.55248 i,0.38227,-1.6647,1.1443,0.0237+1.6373 i,0.16720-0.16022 i,\nnn 0.09765+0.99081 i,0,1.5031,-0.53246-0.03874 i,-1.1259,0.21939-0.79376
   i,-0.26452-1.01231 i,0.80162-0.64935 i,\nnn -0.43261-1.28844 i,-0.13172-0.75084 i,-0.13792-0.80345 i,-0.2954-3.4158 i,0.75140-0.93916 i,-0.15858-0.75731 i,\nnn 0.61414+0.02289 i,0.17208-1.00307
   i,0.5605-2.5543 i,0.77415+0.28436 i,-1.24006-0.20187 i,-0.0941-2.1116 i,\nnn 0.41179-0.61812 i,-0.61289-0.14385 i,0.00420+0.62552 i,0.45087-1.21007 i
\, , \een 
\ben && 0.39164,-1.7022,-0.35103+0.95729 i,0.20259+0.55248 i,0.38227,-1.6647,1.1443,0.0237-1.6373 i,0.16720+0.16022 i,\nnn 0.09765-0.99081 i,0,1.5031,-0.53246+0.03874 i,-1.1259,0.21939+0.79376
   i,-0.26452+1.01231 i,0.80162+0.64935 i,\nnn -0.43261+1.28844 i,-0.13172+0.75084 i,-0.13792+0.80345 i,-0.2954+3.4158 i,0.75140+0.93916 i,-0.15858+0.75731 i,\nnn 0.61414-0.02289 i,0.17208+1.00307
   i,0.5605+2.5543 i,0.77415-0.28436 i,-1.24006+0.20187 i,-0.0941+2.1116 i,\nnn 0.41179+0.61812 i,-0.61289+0.14385 i,0.00420-0.62552 i,0.45087+1.21007 i
\, , \een    
\ben && 0.39164,-1.7022,0.35103+0.95729 i,-0.20259+0.55248 i,0.38227,-1.6647,1.1443,-0.0237-1.6373 i,0.16720-0.16022 i,\nnn -0.09765-0.99081 i,0,1.5031,0.53246+0.03874 i,-1.1259,0.21939-0.79376
   i,0.26452+1.01231 i,0.80162-0.64935 i,\nnn 0.43261+1.28844 i,0.13172+0.75084 i,0.13792+0.80345 i,0.2954+3.4158 i,-0.75140+0.93916 i,-0.15858-0.75731 i,\nnn 0.61414+0.02289 i,-0.17208+1.00307
   i,-0.5605+2.5543 i,0.77415+0.28436 i,-1.24006-0.20187 i,0.0941+2.1116 i,\nnn -0.41179+0.61812 i,-0.61289-0.14385 i,0.00420+0.62552 i,-0.45087+1.21007 i
\, , \een 
\ben && 0.39164,-1.7022,0.35103-0.95729 i,-0.20259-0.55248 i,0.38227,-1.6647,1.1443,-0.0237+1.6373 i,0.16720+0.16022 i,\nnn -0.09765+0.99081 i,0,1.5031,0.53246-0.03874 i,-1.1259,0.21939+0.79376
   i,0.26452-1.01231 i,0.80162+0.64935 i,\nnn 0.43261-1.28844 i,0.13172-0.75084 i,0.13792-0.80345 i,0.2954-3.4158 i,-0.75140-0.93916 i,-0.15858+0.75731 i,\nnn 0.61414-0.02289 i,-0.17208-1.00307
   i,-0.5605-2.5543 i,0.77415-0.28436 i,-1.24006+0.20187 i,0.0941-2.1116 i,\nnn -0.41179-0.61812 i,-0.61289+0.14385 i,0.00420-0.62552 i,-0.45087-1.21007 i
\, , \een   
\ben && 0.81486,-0.81813,0.79755-0.60073 i,-0.47999-0.36154 i,0.78430,-0.81138,0.41009,-0.44206-1.09721 i,-0.34785+0.02123 i,\nnn 0.41308-1.00557 i,0,1.4888,-0.22666+0.09476 i,-1.1367,-0.69448+0.34912
   i,-0.45294+1.25709 i,0.36740+0.11056 i,\nnn 0.27722+0.88306 i,-0.2915+2.1764 i,-0.3057+2.3029 i,-1.7000-0.4665 i,1.3091+0.9847 i,0.13999+0.46494 i,\nnn 0.92298+0.19622 i,0.90835+1.06028
   i,0.4322+1.8622 i,0.27019-0.27159 i,-0.59122+0.18392 i,0.00432+1.26521 i,\nnn 0.96442+0.86007 i,-0.64767-0.00816 i,-0.11705-0.57289 i,1.1066+1.0690 i
\, , \een 
\ben && 0.81486,-0.81813,0.79755+0.60073 i,-0.47999+0.36154 i,0.78430,-0.81138,0.41009,-0.44206+1.09721 i,-0.34785-0.02123 i,\nnn 0.41308+1.00557 i,0,1.4888,-0.22666-0.09476 i,-1.1367,-0.69448-0.34912
   i,-0.45294-1.25709 i,0.36740-0.11056 i,\nnn 0.27722-0.88306 i,-0.2915-2.1764 i,-0.3057-2.3029 i,-1.7000+0.4665 i,1.3091-0.9847 i,0.13999-0.46494 i,\nnn 0.92298-0.19622 i,0.90835-1.06028
   i,0.4322-1.8622 i,0.27019+0.27159 i,-0.59122-0.18392 i,0.00432-1.26521 i,\nnn 0.96442-0.86007 i,-0.64767+0.00816 i,-0.11705+0.57289 i,1.1066-1.0690 i
\, , \een   
\ben && 0.81486,-0.81813,-0.79755+0.60073 i,0.47999+0.36154 i,0.78430,-0.81138,0.41009,0.44206+1.09721 i,-0.34785+0.02123 i,\nnn -0.41308+1.00557 i,0,1.4888,0.22666-0.09476 i,-1.1367,-0.69448+0.34912
   i,0.45294-1.25709 i,0.36740+0.11056 i,\nnn -0.27722-0.88306 i,0.2915-2.1764 i,0.3057-2.3029 i,1.7000+0.4665 i,-1.3091-0.9847 i,0.13999+0.46494 i,\nnn 0.92298+0.19622 i,-0.90835-1.06028
   i,-0.4322-1.8622 i,0.27019-0.27159 i,-0.59122+0.18392 i,-0.00432-1.26521 i,\nnn -0.96442-0.86007 i,-0.64767-0.00816 i,-0.11705-0.57289 i,-1.1066-1.0690 i
\, , \een
\ben && 0.81486,-0.81813,-0.79755-0.60073 i,0.47999-0.36154 i,0.78430,-0.81138,0.41009,0.44206-1.09721 i,-0.34785-0.02123 i,\nnn -0.41308-1.00557 i,0,1.4888,0.22666+0.09476 i,-1.1367,-0.69448-0.34912
   i,0.45294+1.25709 i,0.36740-0.11056 i,\nnn -0.27722+0.88306 i,0.2915+2.1764 i,0.3057+2.3029 i,1.7000-0.4665 i,-1.3091+0.9847 i,0.13999-0.46494 i,\nnn 0.92298-0.19622 i,-0.90835+1.06028
   i,-0.4322+1.8622 i,0.27019+0.27159 i,-0.59122-0.18392 i,-0.00432+1.26521 i,\nnn -0.96442+0.86007 i,-0.64767+0.00816 i,-0.11705+0.57289 i,-1.1066+1.0690 i
\, . \een

}

\end{document}